\documentclass[journal=jacsat,manuscript=article,layout=twocolumn]{achemso}
\usepackage{graphicx}
\usepackage{dcolumn}
\usepackage{bm}
\usepackage{epstopdf}
\usepackage{tabularx}
\usepackage{booktabs}
\let\oldmaketitle\maketitle
\let\maketitle\relax
\setkeys{acs}{articletitle = true, email = false}
\author{Lokanath Patra}
\affiliation[Central University of Tamil Nadu]
{Department of Physics, Central University of Tamil Nadu, Thiruvarur, Tamil Nadu, 610005, India}
\alsoaffiliation[Central University of Tamil Nadu]
{Simulation Center for Atomic and Nanoscale MATerials, Central University of Tamil Nadu, Thiruvarur, Tamil Nadu, 610005, India}
\author{Ravindran Vidya}
\affiliation
{Department of Medical Physics, Anna University, Chennai, 600025, India}
\author{Helmer Fjellv{\aa}g}
\affiliation[University of Oslo]
{Center for Materials Science and Nanotechnology and Department of Chemistry, University of Oslo, Box 1033, Blindern, N$-$0315, Oslo, Norway}
\author{Ponniah Ravindran*}
\affiliation[Central University of Tamil Nadu]
{Department of Physics, Central University of Tamil Nadu, Thiruvarur, Tamil Nadu, 610005, India}
\alsoaffiliation[Central University of Tamil Nadu]
{Simulation Center for Atomic and Nanoscale MATerials, Central University of Tamil Nadu, Thiruvarur, Tamil Nadu, 610005, India}
\title{Giant magnetoelectric coupling in multiferroic PbTi$_{1-x}$V$_x$O$_{3}$ from density functional calculations}
\begin{document}

\twocolumn
[
\begin{@twocolumnfalse}
\oldmaketitle
\begin{abstract}
The giant magnetoelectric coupling is a very rare phenomenon which has gained a lot of attention for the past few decades because of fundamental interest as well as practical applications. Here, we have successfully achieved the giant magnetoelectric coupling in PbTi$_{1-x}$V$_x$O$_{3}$ ( $x=$ 0$-$1) with the help of a series of generalized$-$gradient$-$corrected (GGA), GGA including on$-$site coulomb repulsion ($U$) corrected spin polarized calculations based on accurate density functional theory. Our total energy calculations show that PbTi$_{1-x}$V$_x$O$_{3}$ stabilizes in $C-$type antiferromagnetic ground state for  $x>$ 0.123. With the substitution of V into PbTiO$_3$, the tetragonal distortion is highly enhanced accompanied by a linear increase in polarization. In addition, our band structure analysis shows that for lower $x$ values, the tendency to form 2D magnetism of PbTi$_{1-x}$V$_x$O$_{3}$ decreases. The orbital magnetic polarization was calculated with self$-$consistent field method by including orbital polarization correction in the calculation as well as from the computed X$-$ray magnetic dichroism spectra. A non$-$magnetic metallic ground state is observed for the paraelectric phase for V concentration ($x$) = 1 competing with a volume change of 10\% showing a large magnetovolume effect. Our orbital projected DOS as well as orbital ordering analysis suggest that the orbital ordering plays a major role in the magnetic to non-magnetic transition when going from ferroelectric to paraelectric phase. The calculated magnetic anisotropic energy shows that the direction \lbrack110\rbrack is the easy axis of magnetization for $x$ = 1 composition. The partial polarization analysis shows that the Ti/V$-$O hybridization majorly contribute towards the total electrical polarization. The present study adds a new series of compounds to the magnetoelectric family with rarely existing giant coupling between electric and magnetic order parameters. These results show that such kind of materials can be used for novel practical applications where one can change the magnetic properties drastically (magnetic to non$-$magnetic as shown here) with external electric field and vice$-$versa.
\end{abstract}
\end{@twocolumnfalse}
]

\clearpage

\section{Introduction}
\label{sec:intro} The promising coupling between electric and magnetic order parameters and the potential to manipulate one by the application of the other has created much attention in the past few decades. Research on materials with such coupling has grown interest because of their wide range of application in multifunctional devices.  \cite{fiebig2005revival,spaldin2005renaissance,cheong2007multiferroics,ramesh2007multiferroics,bibes2008multiferroics} The ultimate goal of these research is to obtain single phase multiferroics with strong coupling between ferroelectric and magnetic order parameters at room temperature. There has been a considerable recent interest in developing lone pair based magnetoelectrics because of their high value of electrical polarization. Examples for such composites include BiFeO$_3 -$LaFeO$_3$\cite{gonzalez2012first}, BiFeO$_3 -$SrTiO$_3$\cite{ma2011enhanced}, BiFeO$_3$PbTiO$_3$\cite{zhu2008structural}, BiFeO$_3$--BiCoO$_3$\cite{dieguez2011first}, LaFeO$_3-$ PbTiO$_3$\cite{singh2008magnetization}, BiFeO$_3 -$BaTiO$_3$\cite{hang2012dielectric}, BiCoO$_3 -$BaTiO$_3$\cite{patra2018metamagnetism} etc. PbTiO$_3$ (PTO) is a perovskite ferrolectric material with a Curie temperature of 490$^\circ$ C and a large tetragonal distortion with $c/a$ = 1.06\cite{chen2015negative} at room temperature. But it is non--magnetic due to the absence of $d$ electrons. The Ti atom can be substituted with some magnetic ions as a result of which the new compound can be expected to produce both magnetic and ferrolectric behavior. PTO based magnetoelectrics could be much more interesting because of the existence of two types of mechanisms for ferroelectricity i.e. lone pair electrons from Pb$^{2+}$ and $d^0-$ness from Ti$^{4+}$. So, here we examine the magnetoelectric properties of a PTO based multiferroic series i.e. PbTi$_{1-x}$V$_x$O$_{3}$ ( $x$ = 0, 0.25, 0.33, 0.50, 0.67, 0.75, 1) where the 6$s$ electrons of Pb$^{2+}$ and the $d^0$--ness of Ti$^{4+}$ stabilizes the ferroelectricity and the V induces magnetism.
\par
On the other hand, PbVO$_3$ (PVO) , which is a member of this series ($x=$ 1) has stirred a lot of interest in last few years as a strong candidate for multiferroic oxide due to its large electric polarization. Though it is isostructural with PTO, it shows  large tetragonal distortion ($c/a$ = 1.229). Isolated layers of corner shared VO$_5$ pyramids form a layer type perovskite structure with a space group $P4mm$ \cite{shpanchenko2004synthesis}. PVO has been proposed to have an antiferromagnetic ordering and a ferroelectric polarization as large as 152 $\mu$C/cm$^{2}$ due to the presence of large structural distortions\cite{okos2014synthesis,shpanchenko2004synthesis,belik2005crystallographic,uratani2009first,singh2006electronic}. To this point, however, the true magnetic structure and the multifunctional nature of this material is controversial to researchers. Shpanchenko \textit{et al.}\cite{shpanchenko2004synthesis} found no long-range magnetic ordering with their neutron powder diffraction measurements down to 1.5\,K. Belik \textit{et al.}\cite{belik2005crystallographic} proposed that PVO is a two$-$dimensional spin half square lattice strongly frustrated antiferromagnet due to the antiferromagnetic interactions of the next$-$nearest$-$neighbours. With the help of magnetic susceptibility and specific heat measurements as well as band$-$structure calculations, Tsirlin\cite{tsirlin2008frustrated} confirmed that the S = 1/2 square lattice of Vanadium 4+ ions in PVO is strongly frustrated due to the next$-$nearest$-$neighbor antiferromagnetic interactions and no long-range magnetic ordering was found down to 1.8\,K. Due to the presence of defects or ferromagnetic impurities in the sample, it has been very difficult to clarify the intrinsic magnetic property in PVO. Oka \textit{et al.}\cite{oka2008magnetic} investigated the magnetic properties of PVO by preparing a multidomain single$-$crystal without any magnetic impurity. The broad maximum centered around 180\,K in the temperature dependent magnetization curve indicates the presence of two$-$dimensional antiferromagnetism. Muon spin rotation ($\mu$SR) measurement displayed the presence of a long-range order below 43\,K.   The epitaxial thin films of PVO has been grown\cite{martin2007growth}  using pulsed laser deposition and this brings a step forward to synthesis multiferroic materials outside of high$-$temperature and high pressure techniques to realize devices with multifunctionalities. In another study on PVO thin films by Kumar \textit{et al.}\cite{kumar2007polar}, a transition from a ferroelectric only state to a ferroelectric and magnetic state was determined below 100$-$130\,K using second$-$harmonic generation and X$-$ray linear dichroism. Experimental electron energy loss spectroscopy (EELS) investigation on V$-$L edge show that V in the PVO thin films are in the V$^{4+}$ state resulting in a $d^{1}$ state\cite{chi2007studying}.
\par
PVO was also investigated computationally by different researchers. First principle calculations were made by Uratani \textit{et al.}\cite{uratani2009first} for PVO along with BiCoO$_3$. They found the easy axes of spin are different: [110] in PVO and [001] in BiCoO$_3$ even though both have similar crystal structure. A spin spiral structure was predicted by Solovyev\cite{solovyev2012magnetic} to analyze the absence of long$-$range magnetic ordering in PVO. Calculations along with experiments were performed by Parveen \textit{et al.}\cite{parveen2012thermal} in 2012 to study the thermal properties of PVO and the results confirmed the observations made by Tsirlin \textit{et al.}\cite{tsirlin2008frustrated}. Ming\textit{et al.}\cite{ming2010first} made a comparative study of  the structural, electronic, magnetic and phase transition properties in which various exchange$-$correlation (XC) functionals were used and found that PVO is a 2D $C-$AF where the $d^1$ electron of the V$^{4+}$ ion occupies the $d_{xy}$ orbital. A ferroelectric to paraelectric phase transition at 1.75 GPa was also noticed. Zhou \textit{et al.}\cite{zhou2012structural} revisited the structural transitions in PVO  with the help of a series of X$-$ray diffraction (XRD) measurements and first principle calculations. They found that the $C-$AF insulating and NM metallic states are the ground states for tetragonal and cubic phases, respectively. They have also noticed a noncentrosymmetric tetragonal to centrosymmetric cubic perovskite structural phase transition occurs between a pressure range of 2.7$-$6.4 GPa. 
\par
Milo{\v{s}}evi{\'c} \textit{et al.}\cite{milovsevic2013ab} performed \textit{ab initio} calculations to study the electronic structure and optical properties of PVO (and BiCoO$_3$). With the help of first principle DFT calculations Xing \textit{et al.}\cite{ming2015first} showed a first order tetragonal to cubic phase transition with a volume collapse of 11.6\% under uniaxial pressure of 1.2 GPa accompanied by a $C-$AF insulator to a NM metal. Kadiri \textit{et al.}\cite{kadiri2017calculated} studied the magnetic properties of PVO with the help of first principle calculations and Monte Carlo simulations and they determined PVO as an S = 1/2 antiferromagnet with a N\'eel temperature $T_N$ = 182\,K. Recently, Oka \textit{et al.}\cite{oka2018experimental} examined the cubic phase of PVO under high pressure on experimental and theoretical bases. They determined the transition pressure to be 3 GPa and shown that above which the semiconductor$-$to$-$metal transition associated with the structural transition (tetragonal$-$to$-$cubic) occurs.
\par
There are some reports about the transition metal substitution at the V site of PVO. Tsuchiya \textit{et al.}\cite{tsuchiya2009high} synthesized PbFe$_{0.5}$V$_{0.5}$O$_3$ under high pressure and found the crystal structure as a tetragonal perovskite with $c/a$ = 1.18. The magnetic study revealed that the compound is an antiferromagnet and the electrical polarization is estimated to be as large as 88 $\mu$Ccm$^{-2}$. Ar\'evalo \textit{et al.}\cite{arevalo2011structural} studied Ti and Cr substitutions at V site in PVO. A tetragonal to cubic transition was observed as the Cr substitution level reaches 0.4 where as the Ti substitution preserves the $P4mm$ symmetry. The interesting result observed by Ar\'evalo \textit{et al.}\cite{arevalo2011structural} was the temperature induced phase transition from tetragonal to cubic in PbTi$_{0.8}$V$_{0.2}$O$_3$ at 730\,K. But PVO decomposes to Pb$_2$V$_2$O$_7$ at 570\,K before reaching its transition temperature.\cite{okos2014synthesis} In another V site Ti substituted Pb$M$O$_3$ system (where $M$ stands for V site substituted with Ti), Ricci\cite{ricci2013multiferroicity} predicted multiferroicity in PbTi$_{0.875}$V$_{0.125}$O$_3$ with a ferromagnetic$-$ferroelectric ground state and an electrical polarization of 95 $\mu$C/cm$^{2} $. Recently, Pan \textit{et al.}\cite{pan2017colossal} experimentally studied the Pb(Ti,V)O$_3$ system where the V substitution level varies from 0.1 to 0.6. They have found that the whole composition range the $P4mm$ symmetry was preserved and the tetragonality was abnormally improved (so the spontaneous polarization) with the replacement of Ti with V. Interestingly, they also observed an intrinsic giant volume contraction of ($\sim$3.7\%) for Pb(Ti$_{0.7}$V$_{0.3}$)O$_3$ during the ferroelectric$-$to$-$paraelectric phase transition. These observed interesting findings in the substituted systems motivated us for the present study.
\par
In the present work, we have shown that the V substitution in PTO can induce magnetism which can lead to multiferroicity in the substituted system. Thus our ultimate aim to design magnetoelectric materials from non$-$magnetic system is satisfied. The coupling between the electric and magnetic order parameters is strong due to the presence of two ferroelectric mechanisms (lone pair of Pb$^{2+}$ as well as $d^0-$ness of Ti$^{4+}$). The 2D magnetism arises in PVO due to the $d_{xy}$ type orbital ordering in the ferroelectric ground state. On the other hand, the disorder induced by the substitution may release the frustration in PVO and that can lead to a long-range magnetic ordering. We hope the substitution could reduce the tendency of the system to form a 2D arrangement of V cations and it may lead to form 3D magnetic ordering. 


\section{Computational details}
All the results presented here are obtained from \textit{ab-initio} density functional calculations using the Vienna \textit{Ab initio} Simulation Package (VASP)\cite{kresse1996software} or full potential LAPW method (Wien2k). To obtain the ground state structural parameters for PbTi$_{1-x}$V$_x$O$_{3}$, we have optimized the crystal structure by minimizing the force and stress acting on the system. The generalized gradient approximation (GGA)\cite{perdew1996generalized} in the scheme of Perdew$-$Burke$-$Ernzerhof (PBE)\cite{ernzerhof1999assessment} is employed to treat the exchange$-$correlation as it gives better equilibrium structural parameters than the local density approximation. The ionic positions and the shape of the crystals were relaxed for different unit cell volumes until the energy as well as force convergence criterion were reached i.e. 10$^{-6}$eV per cell and 1 meV\AA$^{-1}$/atom, respectively. As the structural parameters are very sensitive to electrical polarization, a very high plane wave cut$-$off energy of 850 eV\cite{ravindran2006theoretical} was used. A 6x6x6 Monkhorst$-$Pack \textbf{k}$-$point\cite{monkhorst1976special} mesh was used for ferroelectric PTO and similar density of \textbf{k}$-$points was used for all other calculations. To find the ground state magnetic ordering, we have considered non$-$magnetic (NM), ferromagnetic (FM), $A-$type antiferromagnetic ($A-$AF), $C-$type antiferromagnetic ($C-$AF), and $G-$type antiferromagnetic ($G-$AF) configurations\cite{patra2016electronic}. In order to account for the strong electron correlation associated with V 3$d$ electrons on the electronic structure, GGA+$U$ method was used with $U_{eff}$ = 3 eV.The Born effective charges were calculated by the Berry phase method with a 8x8x8 \textbf{k}$-$point mesh per f.u. using the modern theory of polarization\cite{resta1994macroscopic,king1993theory}. We have used periodic supercell approach to simulate the substituted systems i.e. for compositions with 33\%, and 50\% V substitutions 1x1x3, and 1x1x2 supercells were created. Then one Ti atom was replaced with V to mimic the experimental conditions. The supercells are again doubled in each direction to incorporate different magnetic orderings. Similar kind of methodology was used to model other compositions.  
\par
The orbital moments at the transition metal sites were calculated using the orbital polarization correction (SO + OP) method implemented in Wien2k\cite{blaha2001wien2k}. In addition, we have simulated  X$-$ray circular magnetic dichroism (XMCD)\cite{erskine1975calculation} spectra using Wien2k and estimated the site selective spin and orbital magnetic moments of V ions using the XMCD sum rules\cite{carra1993x,thole1992x}. For simulating XMCD, we have used full-potential linearized augmented plane wave (FP-LAPW) method based on density functional theory as implemented in the Wien2k code. The generalized gradient approximation within the PBE scheme was used as the exchange correlation potential. The muffin tin radii for Pb, V and O were chosen as 1.66, 2.49, and 1.50 a.u., respectively. The Brillouin zone integration has been carried out with 5000 k-points. The energy convergence with a convergence criterion of 10$^{-5}$ was obtained. The spin orbit coupling is considered in the calculation. As the ferromagnetic state is a metal, the plasma frequency was first calculated and included in the XMCD calculation. 
\section{Results and Discussions}
\subsection{Structural phase stability}
\begin{figure}[!t]
\includegraphics[scale=0.5]{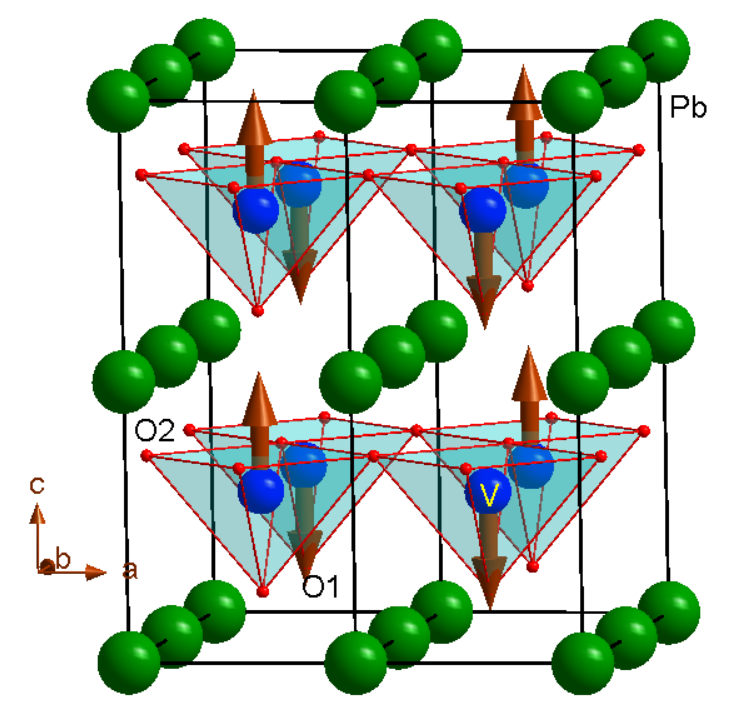}
\caption{ The $C-$type antiferromagnetic structure of PVO. Pb and V atoms are labelled on the illustration as green (big) and blue (medium) spheres, respectively. Oxygen atoms occur at the corners of the VO$_5$ square$-$pyramid and are labelled as red (small) spheres . Two types of oxygen atoms are present in PbVO$_3$ one at apical (labelled as O1) and the other at the planar (labelled as O2) position. The optimized atomic positions are: Pb [0, 0, 0], V [0.5, 0.5, 0.5708 (0.5677)], O1 [0, 0, 0.2128(0.2087)] and O2 [0, 0, 0.6902(0.6919)]. The values in the bracket are from experimental measurements.\cite{belik2005crystallographic}}
\label{afstr}
\end{figure}

PVO is a $C-$type antiferromagnet (See Fig.~\ref{afstr}) due to the presence of a $d$ electron with S = 1/2. In oxides, it can either occur in the middle of an octahedron cage (CaVO$_3$, SrVO$_3$)\cite{nekrasov2005comparative} or it can form a strong vanadyl bond with one O with a short bond length ($\sim$1.55$-$1.60\,\AA)\cite{schindler2000crystal}. PVO is reported to exhibit a tetragonal$-$to$-$cubic phase transition under pressure but no structural transition was observed from 0\,K to its decomposition temperature\cite{belik2005crystallographic}. Moreover, the minimum (apical) V$-$O bond length in the VO$_5$ pyramid in tetragonal PVO is 1.68\,\AA. This relatively longer bond shows that a comparatively weaker vanadyl bond than usual. The average of the four planar V-O bond lengths is 1.99~\AA. In agreement with these findings, the calculated Goldschmidt tolerance factor for PVO also shows a value of 1.036 which usually indicates non-centrosymmetric distortion.
\begin{figure}[!t]
\includegraphics[width=3.2in,height=2.5in]{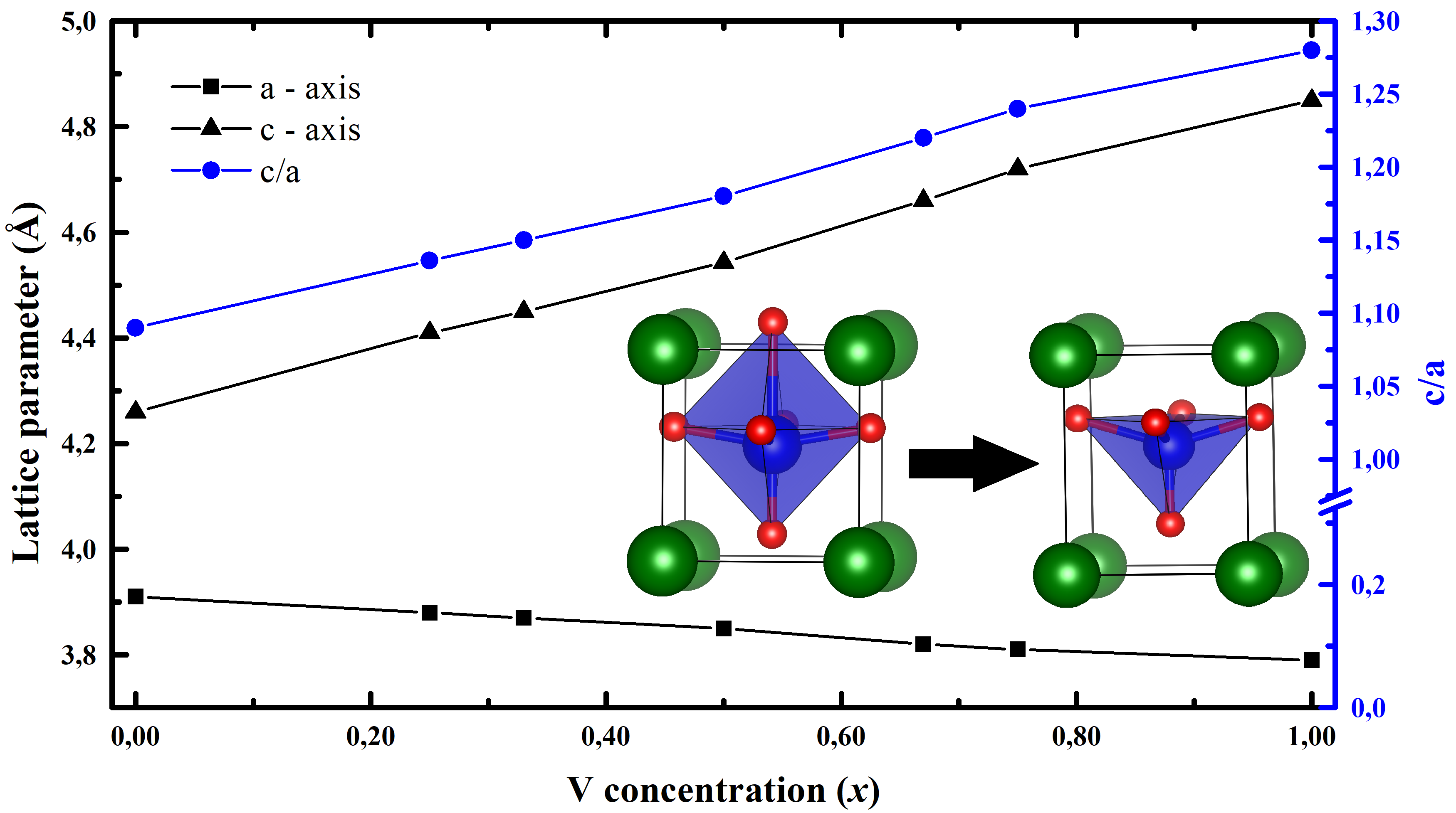}
\caption{ The evolution of lattice parameters in PbTi$_{1-x}$V$_x$O$_{3}$ as a function of $x$. The rapidly increasing $c$ and slowly decreasing $a$ result in unusual enhancement in $c/a$ value. The inset shows the octahedral environment changes into pyramidal one due to increase in tetragonal distortion with increasing $x$.} 
\label{lat_para}
\end{figure}
\par
Here, we have made structural optimizations for both the ferrolectric ($P4mm$) and paraelectric ($Pm\bar{3}m$) phases of PbTi$_{1-x}$V$_x$O$_{3}$ for $x$ = 0, 0.25, 0.33, 0.50, 0.67, 0.75 and 1 compositions. The substitution of V for Ti in PTO doesn't affect the symmetry of the crystal as it retains the tetragonal symmetry for the whole composition range\cite{pan2017colossal}. With increasing V-content ($x$), the lattice parameter $c$ increases almost linearly and $a$ decreases gradually as shown in Fig.~\ref{lat_para}. This results in an unusual enhancement in the tetragonality of the PbTi$_{1-x}$V$_x$O$_{3}$. This can be understood on the basis of ionic radii of Ti$^{4+}$ and V$^{4+}$ ions. The coordination number of Ti$^{4+}$ changes from 6 to 5 whilst increasing concentration of vanadium.  The ionic radius of V$^{4+}$ is 0.53\,\AA; whereas, the ionic radii of Ti$^{4+}$ are 0.61\,\AA~and 0.51\,\AA~for 6 fold and 5 fold co$-$ordinations, respectively\cite{shannon1976revised}. So with the decrease in the ionic radii, the lattice parameter $a$($b$) decreases in order to accommodate the smaller cation. On the other hand, with the increase of V concentration, the chance of vanadyl bond formation increases which leads to an increase in the length of $c-$axis. This is also the reason behind the increase in volume with the increase of $x$. It may be noted that, PTO modified ferroelectrics such as (1-$x$)PTO$-x$Bi(Ni$_{1/2}$Ti$_{1/2}$)O$_3$\cite{choi2005structure} and (1-$x$)PTO$-x$Bi(Mg$_{1/2}$Ti$_{1/2}$)O$_3$\cite{randall2004investigation} usually do not show high value of  $c/a$ ratio. But some of the PTO based compounds i.e. (1-$x$)PTO$-x$BiFeO$_3$\cite{chen2013effectively}, BiInO$_3-$PTO\cite{eitel2001new} and (1-$x$)PTO$-x$Bi(Zn$_{1/2}$Ti$_{1/2}$)O$_3$\cite{suchomel2005enhanced} show abnormally high tetragonality. In the present study, the achieved maximum tetragonality ($c/a$) for a PTO based compound i.e. PbTi$_{0.25}$V$_{0.75}$O$_3$ is 1.24. This value is larger than that of some of the conventional perovskite ferroelectrics such as 0.5PTO$-$0.5BiFeO$_3$ ($c/a$=1.14), 0.85PTO$-$0.15BiInO$_3$ (1.08) and 0.6PTO$-$0.4Bi(Zn$_{1/2}$Ti$_{1/2}$)O$_3$ ($c/a$= 1.11). PTO$-$ based ferroelectric materials with large $c/a$ posses interesting physical properties, such as high value of ferroelectric polarization ({\bf P}$_S$), high Curie temperature ($T_C$), and enhanced negative thermal expansion (NTE). In this study, we have shown that PTO$-$based material with large $c/a$ value possesses strong magnetoelectric coupling as well as giant magnetovolume effect.  

\par
The unit cell volume is sensitive to structural transition in PbTi$_{1-x}$V$_x$O$_{3}$ i.e. it contracts when a ferroelectric to paraelectric transition takes place. For $x$ = 0.25, the volume contracts by 4.2\% during the phase transition. Most interestingly, the volume contraction increases with increase in V content. The volume contraction value increases to 6.3\% for $x$ = 0.50 and 8.4\% for $x$ = 0.67. When $x$ = 1, the volume contracts by a value as large as 10.4\% at the ferroelectric to paraelectric phase transition point. One of the main factors for volume contraction in this series of compounds at the phase transition is the change in bonding between Pb$-$O and Ti/V$-$O which is discussed below. PTO based composites with a high value of volume contraction during phase transition can give rise to interesting properties like high NTE. Moreover, a temperature induced phase transition is observed experimentally for $x$ $\leq$ 0.3\cite{pan2017colossal}. So it can be said that for higher amount of Ti substitution in PbVO$_{3}$, the $T_C$ can be measurable without decomposition.

\par
\begin{figure}[!t]
\includegraphics[scale=0.5]{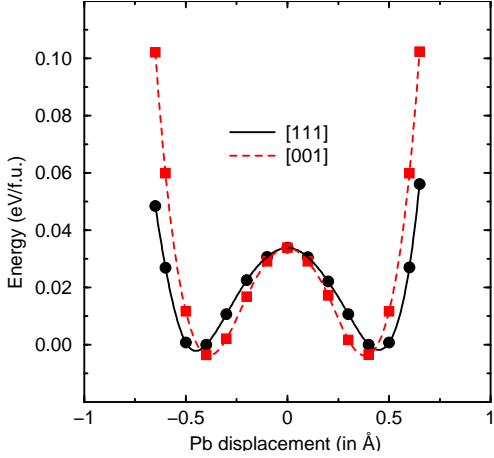}
\caption{(Color online) Total energy as a function of displacement of Pb ion  along [001] and [111] directions for the ferroelectric PVO.  The structural relaxations were consider through selective dynamics.}
\label{fig:dist}
\end{figure}

The cubic structure of PVO can be considered as a special case of ideal tetragonal perovskite with $c/a=$1. The ferroelectric polarization is very sensitive to structural parameters. So, to calculate the $c/a$ ratio for various compositions accurately, we have calculated the total energy as a function of $c/a$. We have performed this calculation with a 1x1x2 supercell which means we should get a minimum energy for $c/a=$2 for a cubic system. In consistent with this, our E $\sim c/a$ graph (See Fig. S1 in supporting information) for PVO ( $x$ = 0) shows a minimum energy for $c/a=$2 for the paraelectric phase; whereas, for the ferroelectric phase, we get a non$-$integer $c/a$. In order to find the possible polarization path, we have calculated the total energy as a function of Pb displacement with respect to V/Ti$-$O polyhedra. Figure \ref{fig:dist} helps us to find the easiest path of the ferroelectric polarization and the role of Pb ions in the tetragonal distortion. We have plotted the displacement of Pb ion with respect to VO$_5$ pyramid along [111] and [001] directions. The energy vs displacement curves for both the directions have double well shapes. It can be seen that the lowest energy for the off$-$center displacements is along the [001] direction. It shows that the ferroelectric properties are greatly effected by the polarizability of Pb. The energy difference between the two directions is very small (about 2\,meV/f.u.) compared to well$-$known multiferroic BiFeO$_3$\cite{ravindran2006theoretical}. So, in principle, the Curie temperature of PVO must be less than that of BiFeO$_3$. 
However, the $T_C$ for PVO has not been estimated yet, as it decomposes before reaching to its paraelectric phase. But the $T_C$ of PbTi$_{0.9}$V$_{0.1}$O$_3$ was estimated to be 823\,K by Zhao Pan \textit{et. al.}~\cite{pan2017colossal} and it showed an increasing trend V concentration (Measured $T_C$ for PbTi$_{0.8}$V$_{0.2}$O$_3$ and PbTi$_{0.7}$V$_{0.3}$O$_3$ are in between 823\,K to 873\,K). Compositions with $x>$ 0.3 decompose before reaching their $T_C$.

\begin{figure}[!t]
\includegraphics[scale=0.3]{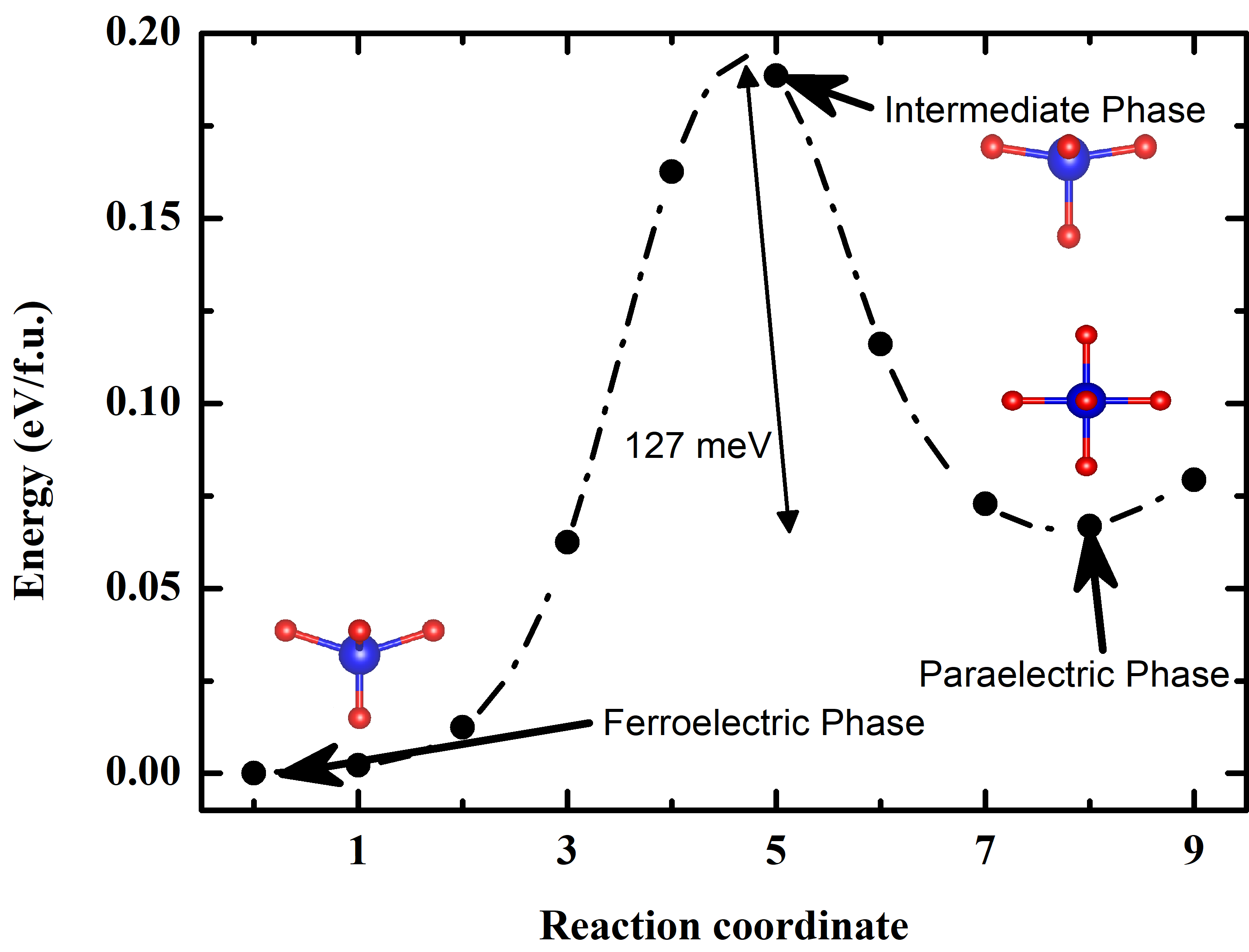}
\caption{ The ferroelectric to paraelectric transition path for PVO obtained from nudged elastic band method. Both volume and shape variation during the transition are considered. The insets show the V and O coordination for the particular images marked by arrows. The atoms colors are same as in Fig. ~\ref{afstr} } 
\label{fig:neb}
\end{figure}

We have also performed nudged elastic band (NEB) calculations in order to find out the energetics of ferroelectric to paraelectric phase transition and to identify a minimum energy path for the phase transition. To calculate the minimum energy path, we have considered the tetragonal phase of PbVO$_3$ as the initial structure whereas the cubic phase as tetragonal structure. Different intermediate images were created and the energies for all the images were calculated considering the volume as sell as shape variation. The energies obtained are plotted versus the images (here mentioned as reaction coordinates).The graph given in Fig. \ref{fig:neb} is resulted from NEB calculations and it shows that the ferroelectric--to--paraelectric transition involves an energy barrier of 127 meV which is higher than the difference between the ground−state total energies of these two phases i.e. $\sim$70 meV. It also shows the V--O coordinated polyhedra. In case of ferroelectric phase, the V--O polyhedron is highly distorted and makes an pyramidal arrangement. The intermediate images along the FE--to--PE transition path with higher energies than the FE phase possess comparatively less distorted polyhedra. Finally, the end product i.e. the paraelectric phase exhibits an undistorted octahedron. The high energy barrier for the phase transition is consistent with the high temperature FE--to--PE phase transition.

\par

\subsection{Chemical Bonding}
The chemical bonding analysis in compounds similar to PVO i.e. BaTiO$_3$ and PTO shows that the ferroelectric instability arises due to the hybridization interaction between Ti 3$d$ and O 2$p$ states\cite{cohen1992origin}. So it is interesting to analyze the chemical bonding between V/Ti and O in PbTi$_{1-x}$V$_x$O$_{3}$ in order to reach the depth of the origin of ferroelectricity. Experimental results and theoretical analysis also show that a significant hybridization is present between Pb and O\cite{kuroiwa2001evidence}. This also contributes to the Born effective charges (BEC) and consequently to the polarization. 
\par
\begin{figure}[!t]
\includegraphics[scale=0.10]{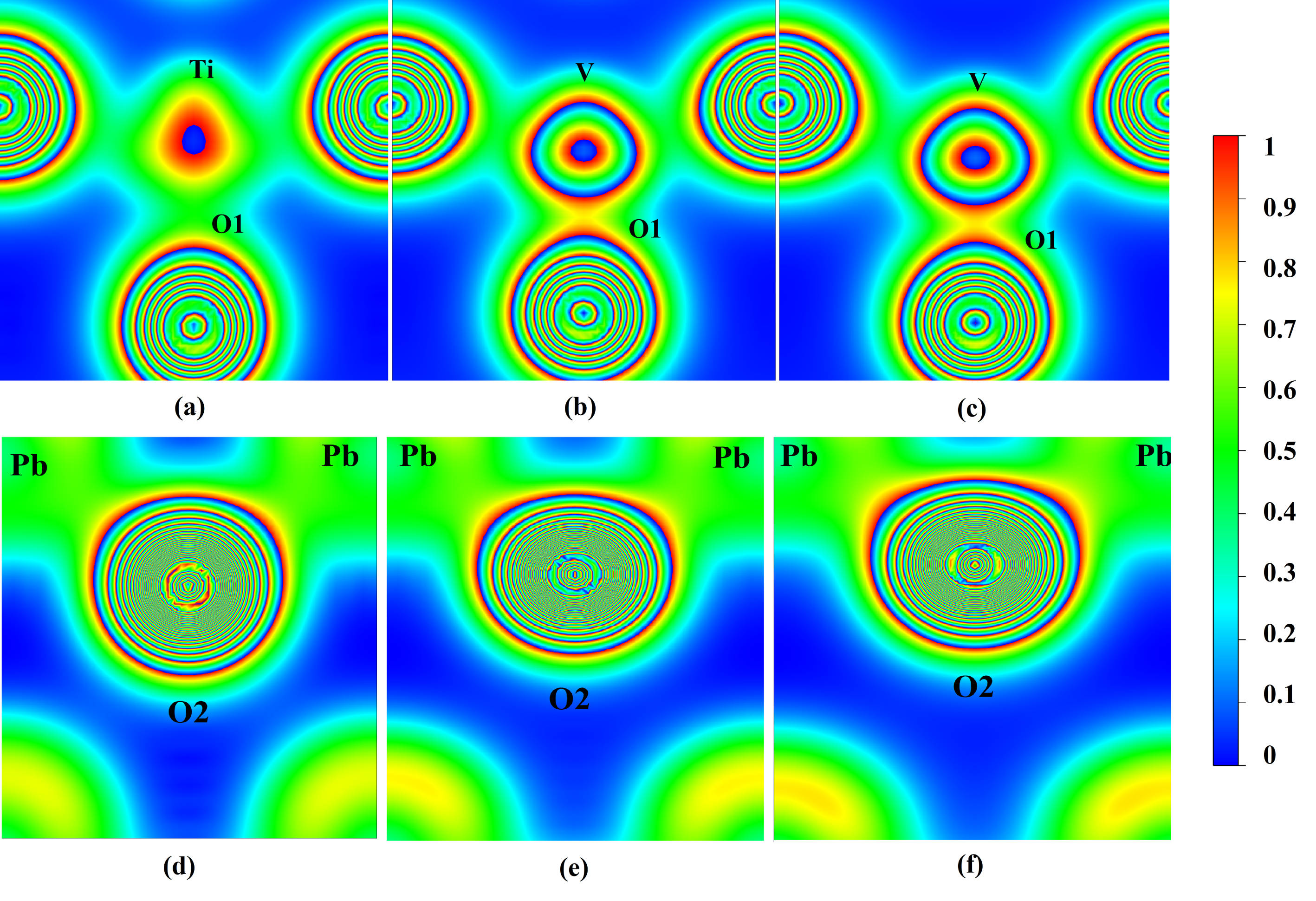}
\caption{\label{fig:charge} Charge density between (a) Ti and O1 in PbTiO$_{3}$, (b) V and O1 in PbTi$_{0.5}$V$_{0.5}$O$_{3}$, (c) V and O1 bond in PbVO$_{3}$, (d) Pb and O2 in PbTiO$_{3}$, (e) in PbTi$_{0.5}$V$_{0.5}$O$_{3}$, and (f) in PbVO$_{3}$.}
\label{charge}
\end{figure}

The charge density plot for $x$ = 0, 0.50, and 1 compositions are given in  Fig. \ref{charge}.  It can be seen from the Fig.\ref{charge} (a), (b), and (c) that the charge density between the $B-$site cation and apical oxygen increases as the V concentration increases. In PVO, the  presence of strong V$-$O1 bond  along with lone pair electrons at the Pb$^{2+}$ site weaken the Pb$-$O1 bond. Due to these, the Pb atom shifts more towards O2 to maintain the charge balance. Hence, the covalent bonding interaction between Pb and O2 shows increasing trend which is evident from Fig.\ref{charge} (d), (e) and (f) as the width of the charge density increases with increase in $x$. These observations indicate that the increase of V concentration in PbTi$_{1-x}$V$_x$O$_{3}$ not only increases the covalent bonding between Ti/V$-$O1 but also that between Pb$-$O2. As a consequence of these, the tetragonal distortion increases with increase in $x$. 

\par
\begin{figure}[!t]
\includegraphics[scale=0.3]{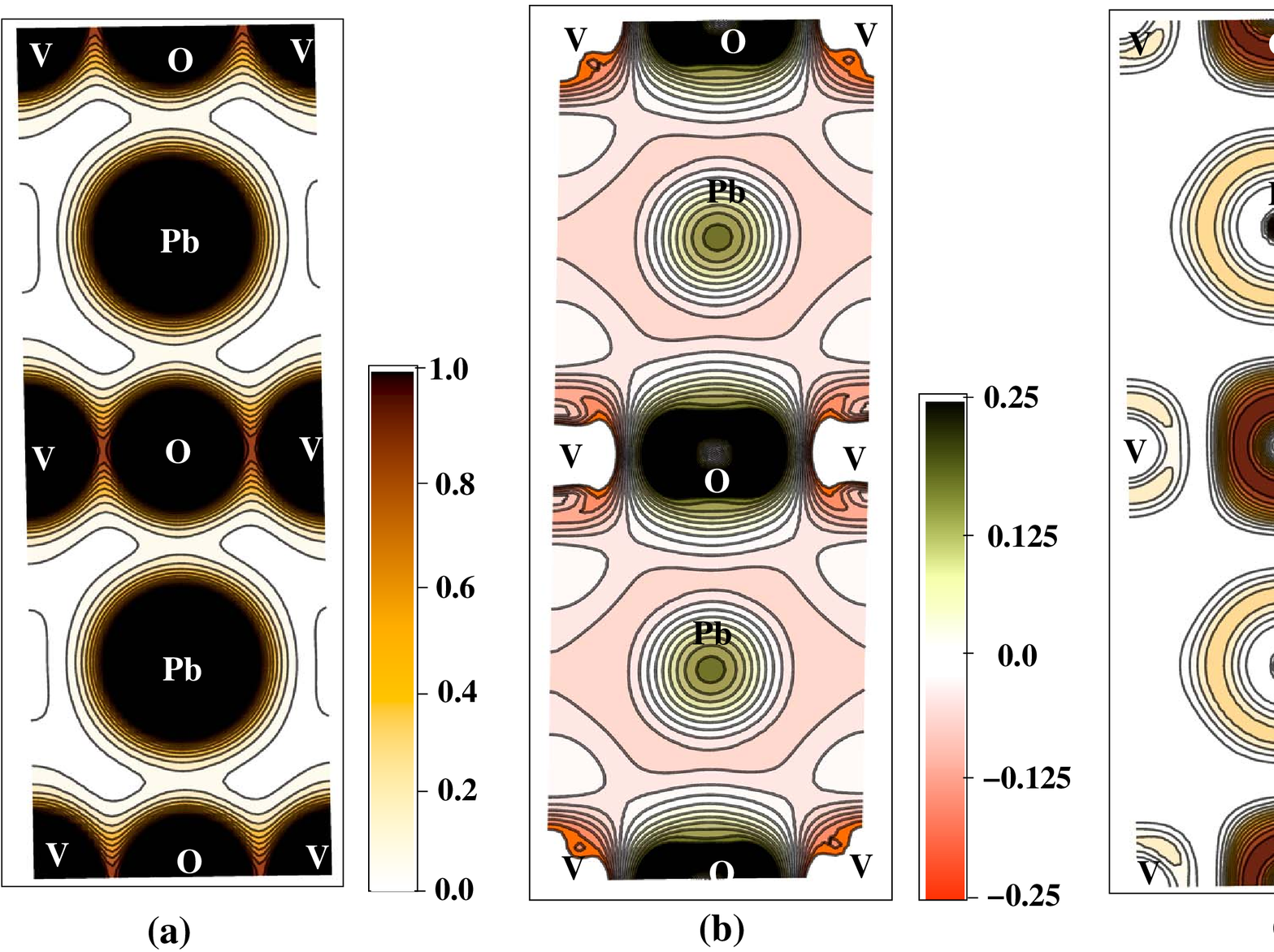}
\newline
\includegraphics[scale=0.3]{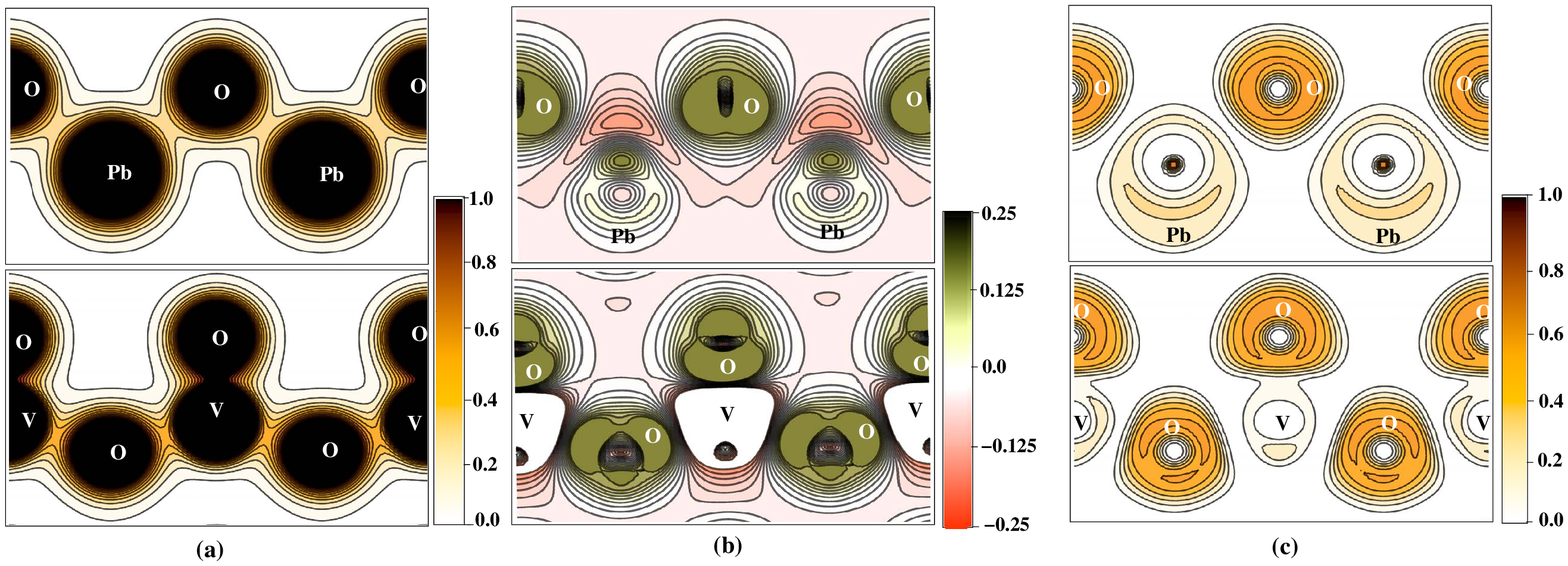}
\caption{(a) Charge density, (b) Charge transfer, and (c) Electron localization function for the paraelectric (upper panel) and ferroelectric (lower panel) structures.}
\label{charge_fp}
\end{figure}

The charge density plots for PVO in cubic and tetragonal phases are given in Fig. \ref{charge_fp}(a) upper and lower panel, respectively. It can be seen that, the charge density distribution at the Pb and O sites are essentially isotropic in nature for the paraelectric phase. On the other hand, finite amount of charge can be seen in between Pb and O in the tetragonal phase showing an anisotropic nature of charge density distribution. So, it can be stated that in cubic phase the Pb-O bond has more ionic character and the tetragonal phase has a mixed iono-covalent bonding character. On the other hand, the bonding between V and O is found to have substantial covalency in  both the cubic and tetragonal phases as we can find noticeable charge density between these atomic sites. But, it may be noted that, the covalent interaction between V and O in tetragonal phase is stronger than that in the cubic case. Due to this, the V atom is shifted from the center of the octahedron stabilizing the ferroelectric state..
\par
Charge transfer distribution shows isotropic nature at the Pb and O sites (Fig.~\ref{charge_fp}(b)) in the cubic phase confirming the ionic bonding  between Pb$-$O. The anisotropic nature of charge transfer distribution between V and O indicates the presence of finite covalent bonding between these atoms in the cubic phase. The charge transfer distribution between V and O as well as Pb and O in tetragonal phase shows anisotropic behavior as shown in Fig. \ref{charge_fp}(b) indicating the presence of  finite covalent bonding between them. It may be noted that, though the covalency between Pb$-$O is weaker than that between V$-$O in the tetragonal phase, it has equal importance in structural and ferroelectric properties. 
 \begin{figure}[!b]
 \includegraphics[scale=0.2]{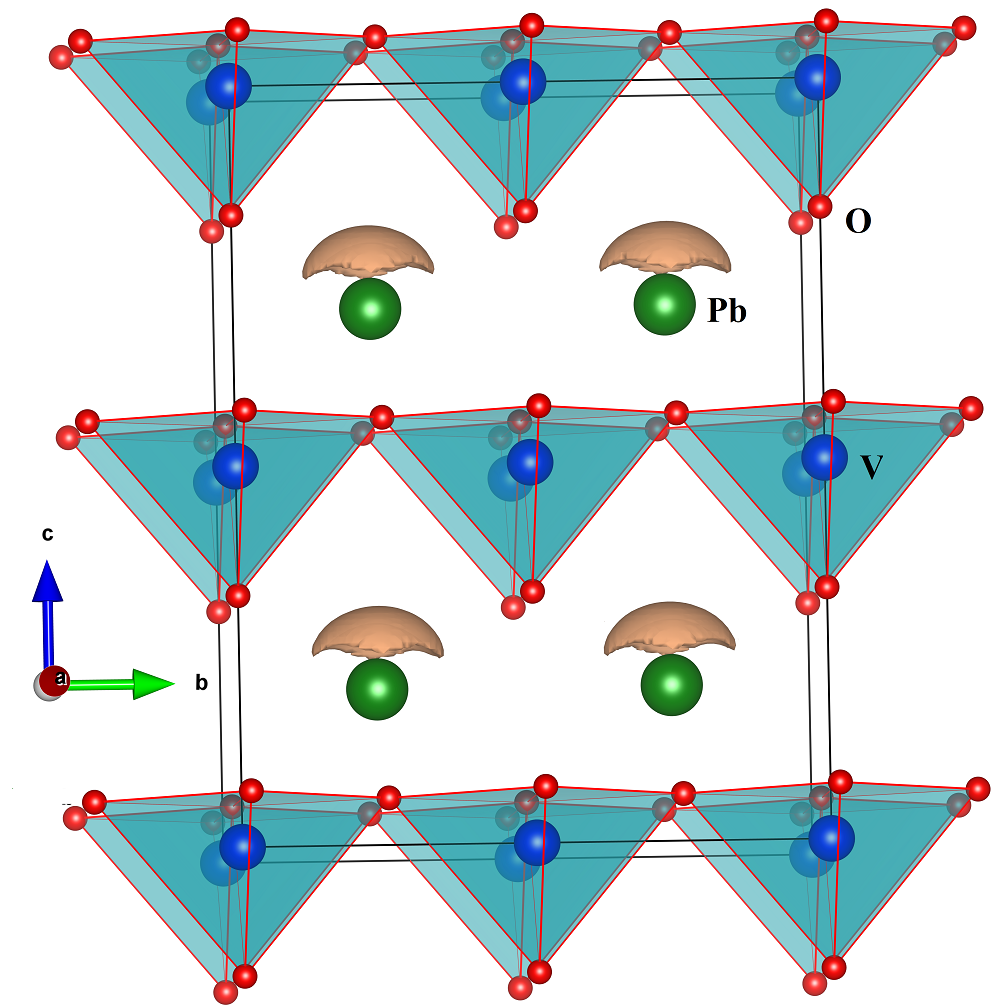}
 \caption{ (Color online) The valence electron localization function isosurface plotted at a value of 0.7 for PVO in the ferroelectric $P4mm$ structure.}
 \label{ELF}
 \end{figure}
\begin{figure*}
\includegraphics[width=2.5in,height=4in]{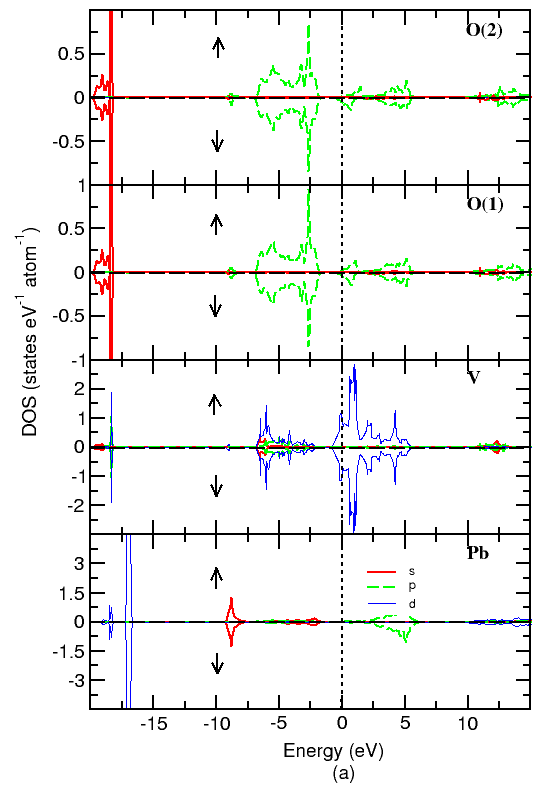}
\includegraphics[width=2.5in,height=4in]{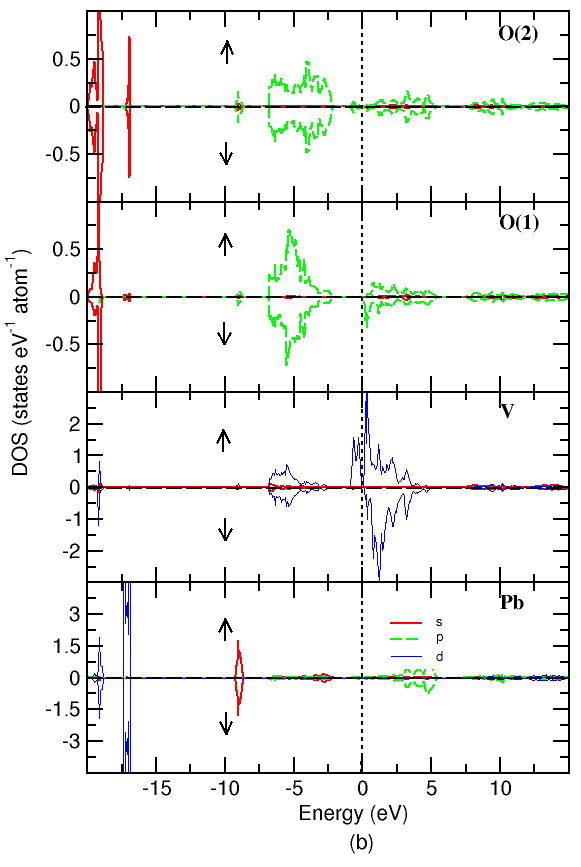}
\caption{\label{fig:bnds}The site and orbital projected density of states (DOS) for PVO for the respective ground state magnetic structures in the  (a) paraelectric and (b) ferroelectric phases. The Fermi level is set to zero.}
\label{partial_dos}
\end{figure*}
\par
Electron localization function (ELF) also provides important information about the bonding interaction between the constituents of a compound.  It can be seen from Fig. \ref{charge_fp}(c) that the ELF is maximum at O sites and minimum at Pb and V sites for both the paraelectric and ferroelectric phases. Finite ELF can be seen in between V and O in the tetragonal phase showing the presence of finite covalent bonding between them. The presence of the stereochemically active lone pair electrons at Pb site is also clearly visible in the ELF plot for the tetragonal case. So the bonding interaction between Pb and O as well as V and O in the ferrolectric tetragonal phase can be concluded as having dominant ionic character with finite covalent bonding. The difference in the bonding behaviour of Ti/V$-$O and Pb$-$O bonds for paraelectric and ferroelectric phases play a crucial role for the large volume contraction during the ferroelectric-to-paraelectric phase transition. We have also plotted the isosurface of the valence electron localization function with an ELF value of 0.7 and given in Fig. \ref{ELF}. This figure shows the lobe like lone pairs of hybridized 6$s$ electrons at the Pb site. The charge density and charge transfer plot show that finite amount of electrons from Pb sites covalently interact with O 2$p$ ensuring that 6s electrons of Pb are not completely chemically inert.

\par
The site and orbital projected DOS for paraelectric and ferroelectric phases for their respective ground state magnetic configurations are given in Fig. \ref{partial_dos} (a) and (b), respectively.  It can be seen that the V 3$d$ states and O 2$p$ states are spread over in the valence band in a range of -7\,eV to Fermi level (E$_F$). The energetically degenerate nature of these two states indicates the presence of strong covalent hybridization between V and O. This hybridization weakens the short range repulsion to lower the energy of the ferroelectric phase. The filled O 2$s$ states form narrow bands around -18\,eV (see Fig. \ref{partial_dos}(b)). Above that, at around -9\,eV, the Pb 6$s$ lone pair electronic states are present with small contributions from 2$p$ electrons. This also confirms that the 6$s$ electrons are stereochemically active and not inert. Though the Pb 6$p$ states are well separated from its 6$s$ states and are mostly present in CB, a small amount of Pb 6$p$ states can be seen which are degenerate with O 2$p$ states confirming the Pb$-$O covalency.  The hybridization between these states enhances the stability of the ferroelectric phase by lowering the total energy of the system. The partial DOS for PbTi$_{0.5}$V$_{0.5}$O$_3$ calculated with GGA$+U$ ($U_{eff}$ = 3 eV) is given as Fig. S2 in the supporting information. Similar kind of DOS distribution can be seen for Pb, V and O atoms in PbTi$_{0.5}$V$_{0.5}$O$_3$ also. The almost empty $d$ states indicates the $d^0-$ness of Ti$^{4+}$. Due to the covalent interaction between Ti and O, the charges redistribute among themselves for which we see very small yet noticeable DOS in Ti valence band. 

\par
\begin{figure}[!t]
 \includegraphics[scale=0.35]{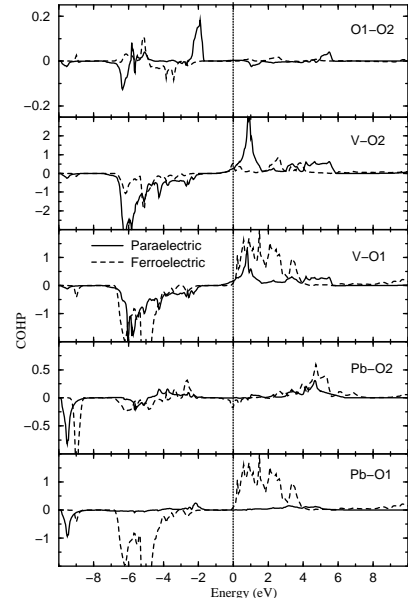}
 \caption{ Crystal orbital Hamilton population (COHP) for PVO in the paraelectric and ferroelectric phases describing Pb$-$O, Pb$-$O2, V$-$O1, V$-$O2, O1$-$O2. O1 and O2 are the planar and apical oxygen atoms, respectively.}
 \label{fig:cohp}
 \end{figure}

The nature and strength of the bonding can also be determined by crystal orbital Hamilton population (COHP) analysis where the negative and positive COHP  values indicate bonding and anti$-$bonding interactions, respectively. We have calculated the COHP among the constituents of PVO (Pb$-$O, Pb$-$O2, V$-$O1, V$-$O2, O1$-$O2) and given in Fig. \ref{fig:cohp}. It can be seen that, all the occupied states for Pb$-$O and V$-$O interactions have bonding states for both paraelectric and ferroelectric phases. The Pb$-$O2 interaction for both the paraelectric and ferroelectric phases are almost the same. But the interaction between Pb$-$O1 in the ferroelectric case has more bonding states which are due to the hybridization between O 2$p$ and Pb 6$s$/6$p$ states. This indicates the presence of stronger covalency between Pb and O1 in case of the ferroelectric phase than the paraelectric one which is also consistent with our charge density and partial DOS analyses. In the ferroelectric phase the V$-$O1 bonding interaction is stronger than V$-$O2 bonding interaction which is due to the off$-$center displacement of V atom towards the O1 atom. 
\begin{figure*}
\includegraphics[scale=0.4]{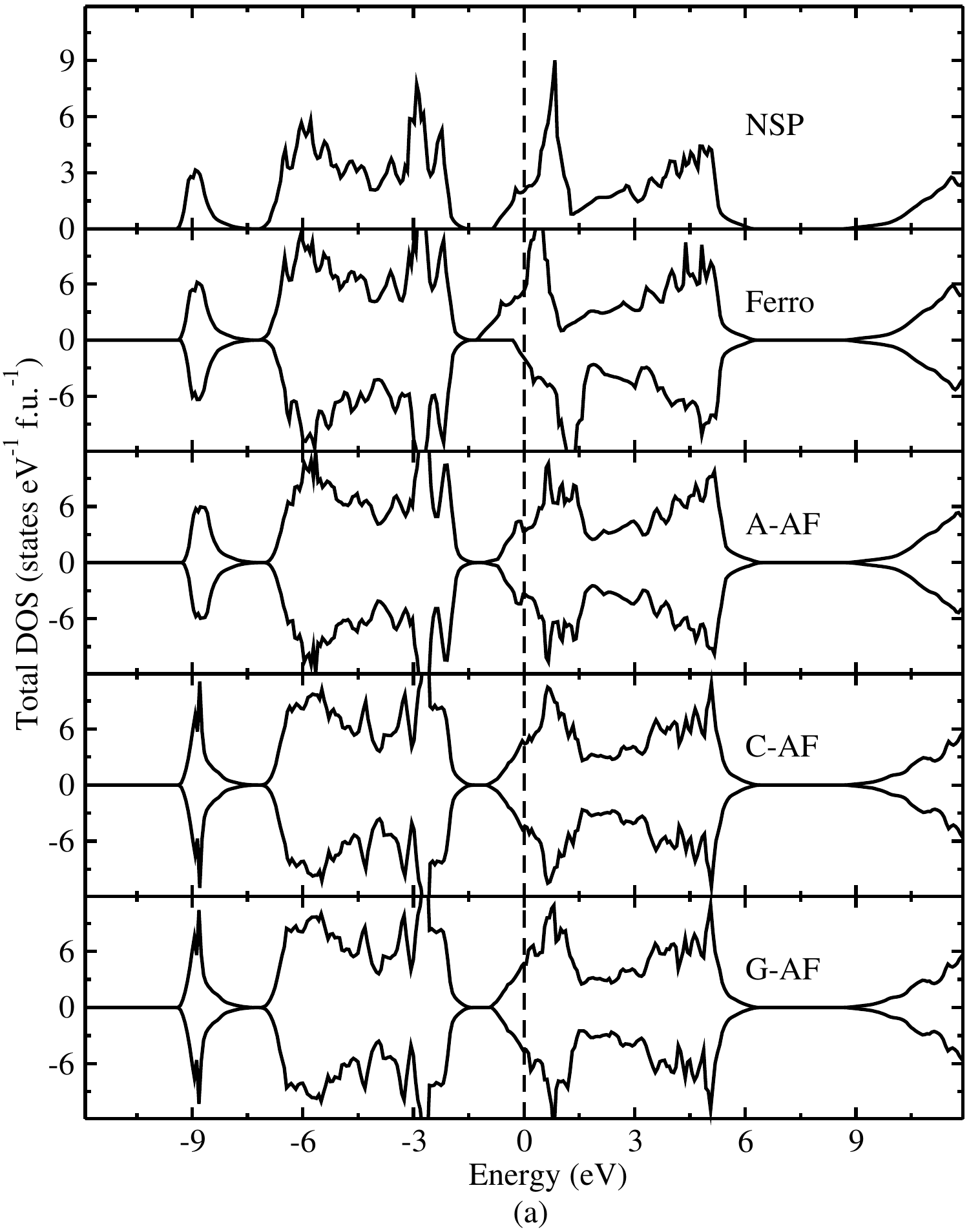}
\includegraphics[scale=0.4]{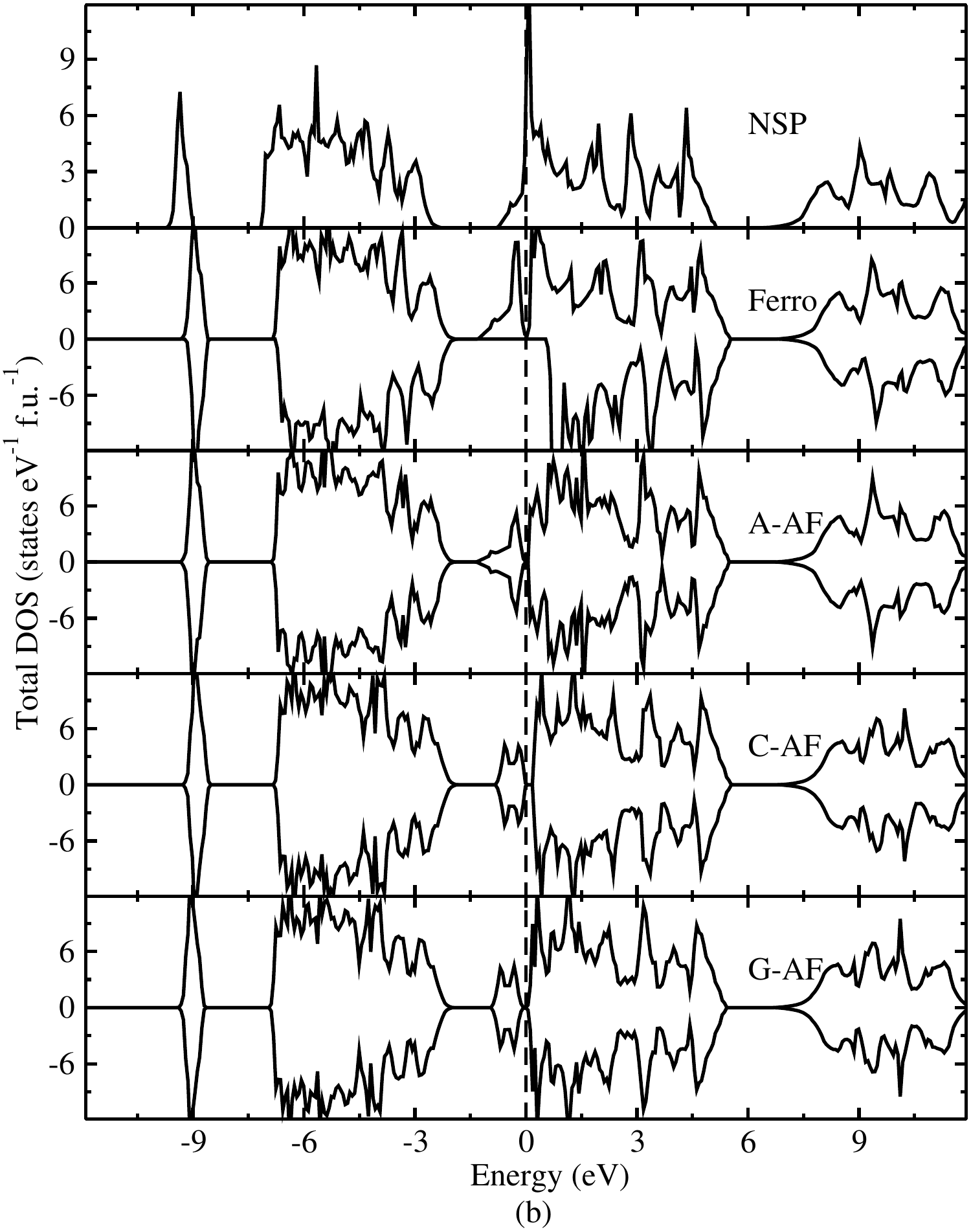}
\caption{Total density of states for PVO in the nonspin polarized (NSP), ferromagnetic, $A-$, $C-$, and $G-$type antiferromagnetic configuration for (a) paraelectric and (b) ferroelectric phases. The Fermi level is set to zero.}
\label{total_dos}
\end{figure*}
\subsection{Electronic Structure}
The total DOS given in Fig. \ref{total_dos} (a) for the paraelectric phase shows a metallic behavior for all magnetic configurations. From the total DOS of the ferroelectric phase (Fig. \ref{total_dos}(b)), it can be seen that the nonmagnetic phase shows a metallic state; whereas, the ferromagnetic case is a half metal with a gap of $\sim$ 0.5\,eV in the down spin channel. But, when the antiferromagnetic ordering is introduced, the exchange potential produced by the exchange interaction pushes the V 3$d$ states towards the lower energy side which results in an insulating state with a narrow gap. So, a combination of spin polarization, magnetic ordering as well as crystal structure play important roles in stabilizing the ferroelectric phase of PVO.  The broad band spreading from -7 to -2\,eV is mainly due to the strong hybridization between V 3$d$  O 2$p$ electrons. The localized states at around -9\,eV are generated from the hybridized Pb 6$s$ lone pair electrons.  After including the correlation effect through the Hubbard $U$ into the Hamiltonian matrix, the states get more localized to give a larger band gap and this localization of the $d$ electrons produces a larger magnetic moment.

\par
The total DOS for $x$ = 0.33, 0.50 and 0.67 compositions were calculated with the GGA+$U$ method and given as Fig. S3 in the supporting information. The band gap value decreases with increase in $x$. As it is known that the bond lengths and lattice parameters have direct impact on band gaps,  the decreasing band gap may be attributed to the increase in the average Ti/V$-$O bond length with $x$ in Ti/V$-$O$_6$ polyhedra (The average Ti$-$O bond length is 2.07~\AA in PTO whereas the V$-$O bond length is 2.11~\AA in case of PVO). The lattice parameter $c$ also increase with $x$. 
\par
\begin{figure*}
\includegraphics[width=3.0in,height=2.5in,angle=270]{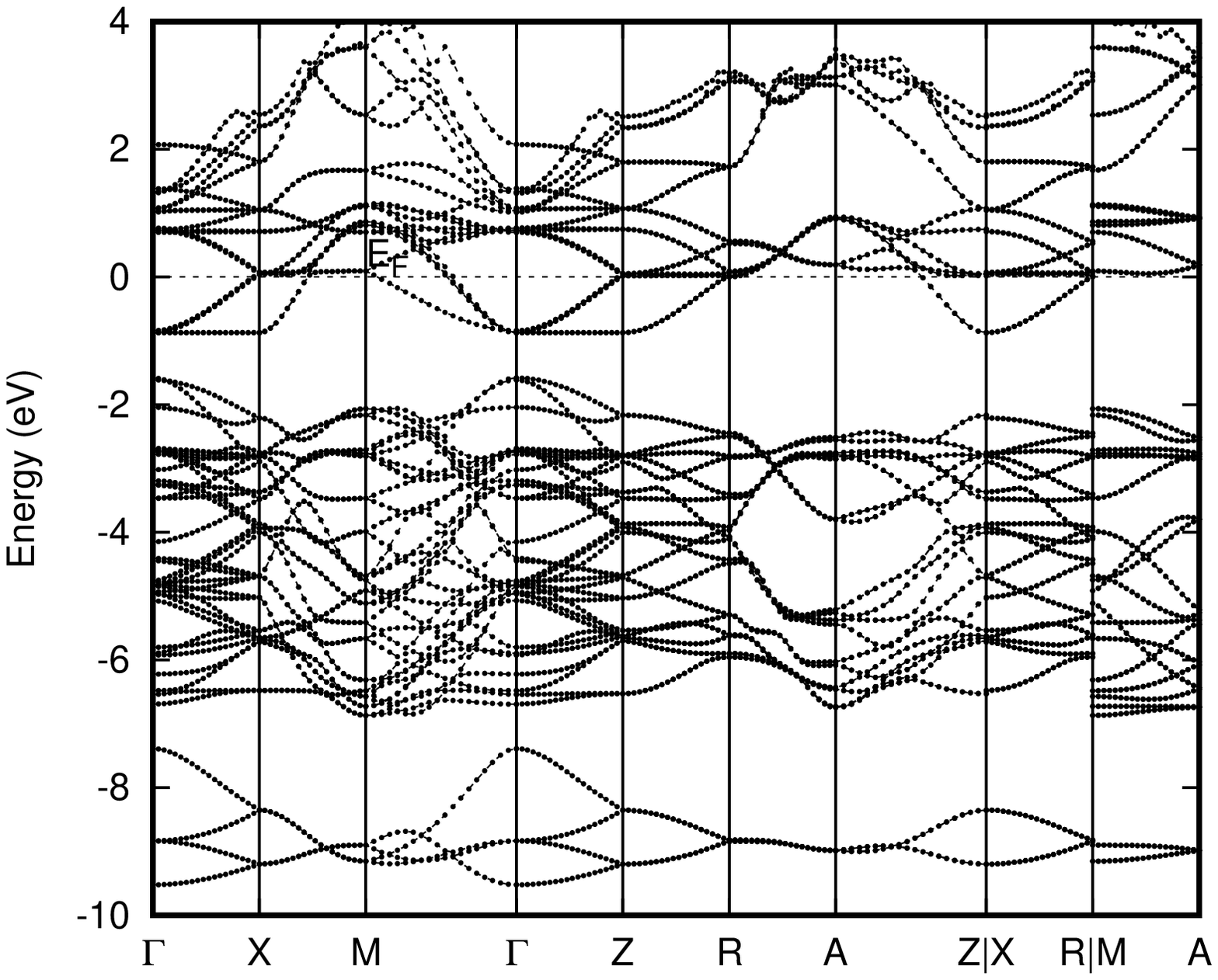}
\includegraphics[width=3.0in,height=2.5in,angle=270]{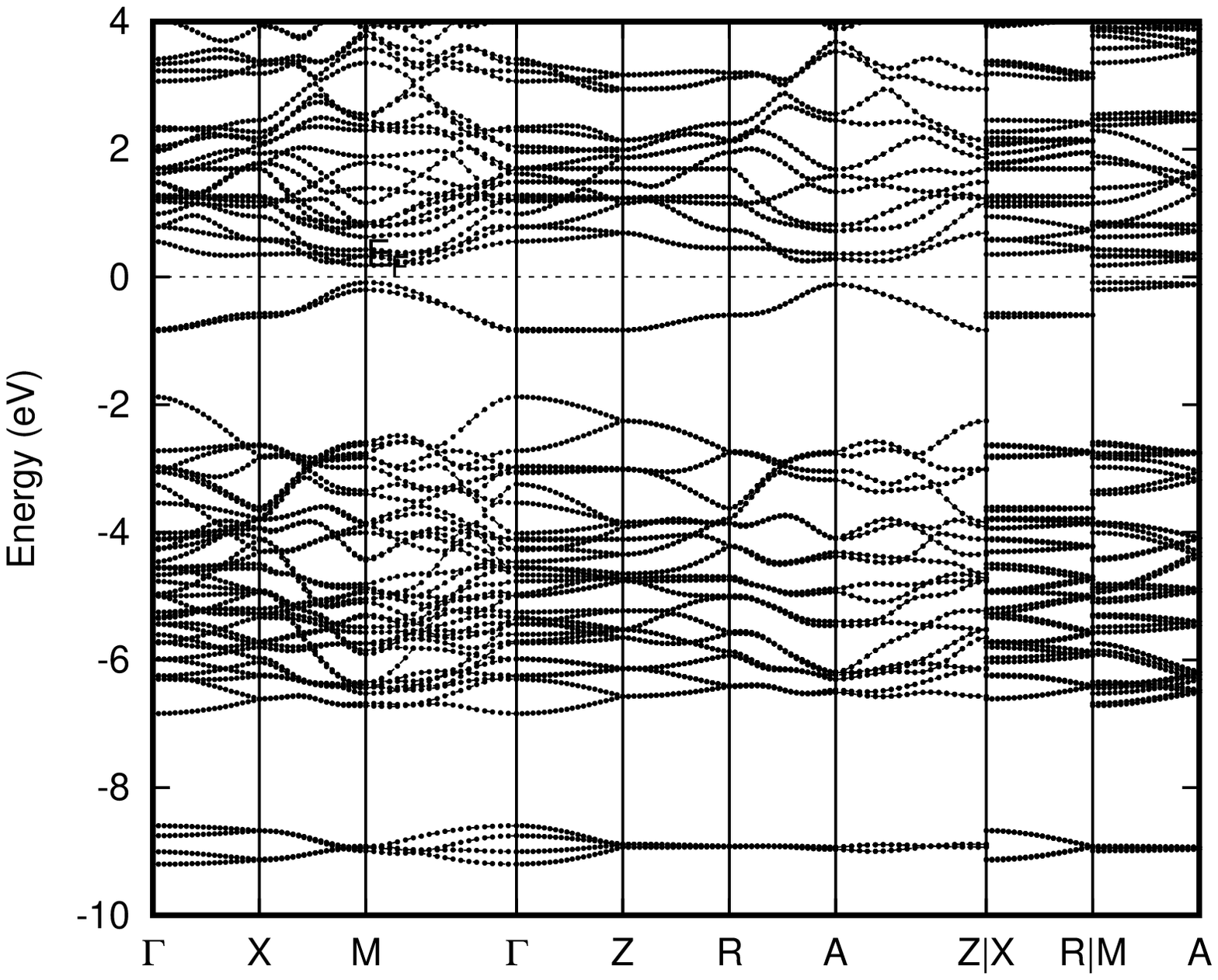}
\caption{Electronic band dispersion for PVO in the (left) paraelectric  and (right) ferroelectric phases. The Fermi level is set to zero.}
\label{bnds}
\end{figure*}

The band structure for both the para$-$ and ferroelectric phases of $x$ = 1 composition are given in Fig. \ref{bnds}. The \textbf{k}$-$path considered in the band structure is given as Fig. S4 in the supporting information. The lowest bands seen around 9\,eV below the E$_F$ are due to the Pb 6$s$ states. It is to be noted that, these bands are broader in the paraelectric phase; whereas, in the ferroelectric phase, they are more localized in a small energy range. This explains the effect of 6$s$ lone pair in stabilizing the ferroelectric state. A manifold of occupied O 2$p$ bands are spread over a range from -7\,eV to -2\,eV. A pair of bands separated from O 2$p$ bands and localized just below $E_F$. These are having V 3$d$ character with no dispersion along $\Gamma-$Z direction and a very weak dispersion along X$-$R as well as M$-$A directions indicating a 2D characteristic in the $ab-$plane\cite{singh2006electronic}. Above these band features a narrow gap with the value of 0.2\,eV is present in the ferroelectric phase. The conduction bands located in the vicinity of CB-edge are  derived from V 3$d$ states, but the bands at higher energy ($\sim$ 3\,eV) have mainly Pb 6$p$ character. We have also calculated the band structure with GGA+$U$ method. The GGA+$U$ band structures for $x$ = 0.5 and $x$ = 1 compositions are given in Fig. S5 and S6 in the supporting information. It can be seen that when the V concentration is less, the vanadyl bond weakens and therefore the $d_{xy}$ band gets dispersed along $\Gamma-$Z, X$-$R and M$-$A directions. This indicates that the tendency of the system to form the 2D magnetism is reduced for lower $x$ values.
\subsection{Magnetic Properties}
We have calculated total energies for nonmagnetic and all the magnetic orderings mentioned above for PbTi$_{1-x}$V$_x$O$_{3}$ in both paraelectric and ferroelectric phases and given in Table S1 in the supporting information. PTO is a well$-$known non$-$magnetic compound. Our total energy calculation reveals that the $C-$AF is the ground state for other compositions. It is to be noted that the $C-$AF and $G-$AF states are almost degenerate (See Table S1 in the supporting information). The present observation of $C-$AF state as a magnetic ground state for PVO is in consistent with experimental observation of a two$-$dimensional $C-$AF phase and also with other theoretical studies\cite{shpanchenko2004synthesis, uratani2009first}. To identify the exact composition where the non$-$magnetic to magnetic transition happens, we have plotted $\Delta$E vs $x$ (where $\Delta E=E_{C-AF}-E_{NM}$) for the ferroelectric phase and given in the supporting information Fig. S7. The energy difference increases linearly as the concentration of V increases due to the increase in localized $d$ electrons density per cell making the antiferromagnetic state more stable. We have extrapolated the $\Delta$E vs $x$ curve and the magnetic transition point is found to be $x\sim$ 0.123. This is in agreement with previous theoretical finding by Ricci \textit{et al.}\cite{ricci2013multiferroicity} using similar calculations that the magnetic ground state stabilize in PbTi$_{1-x}$V$_x$O$_{3}$ for $x$ = 12.5\%.  

\par
The total energy shows a non$-$magnetic ground state for the paraelectric phase of PbTi$_{1-x}$V$_x$O$_{3}$. So the ferroelectric to paraelectric phase transition is associated with a magnetic to non$-$magnetic transition which shows that a strong coupling between the electric and magnetic order parameters for $x>$ 0.123. Now, the large values of volume contraction during the structural transition can also be attributed to the associated magnetic phase transition. So it can be said that, this series of compounds possess not only giant magnetoelectric effect, but also giant magnetovolume effect. It is to be noted that compounds with strong magnetovolume effect can be used to convert magnetic energy into kinetic energy, or vice-versa, and are used to build actuators and sensors. 

\begin{figure}[!b]
\includegraphics[scale=0.4]{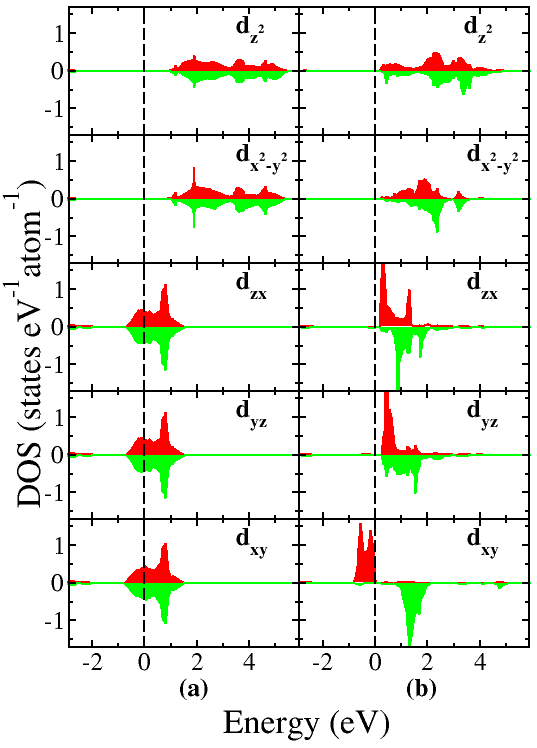}
\caption{The orbital projected density of states for the 3$d$ electrons in the V site of PbVO$_3$ in the  (a) paraelectric and (b) ferroelectric phases for their respective ground state magnetic structures. The Fermi level is set to zero.}
\label{orb}
\end{figure}

\begin{figure*}
\includegraphics[width=3in,height=3in]{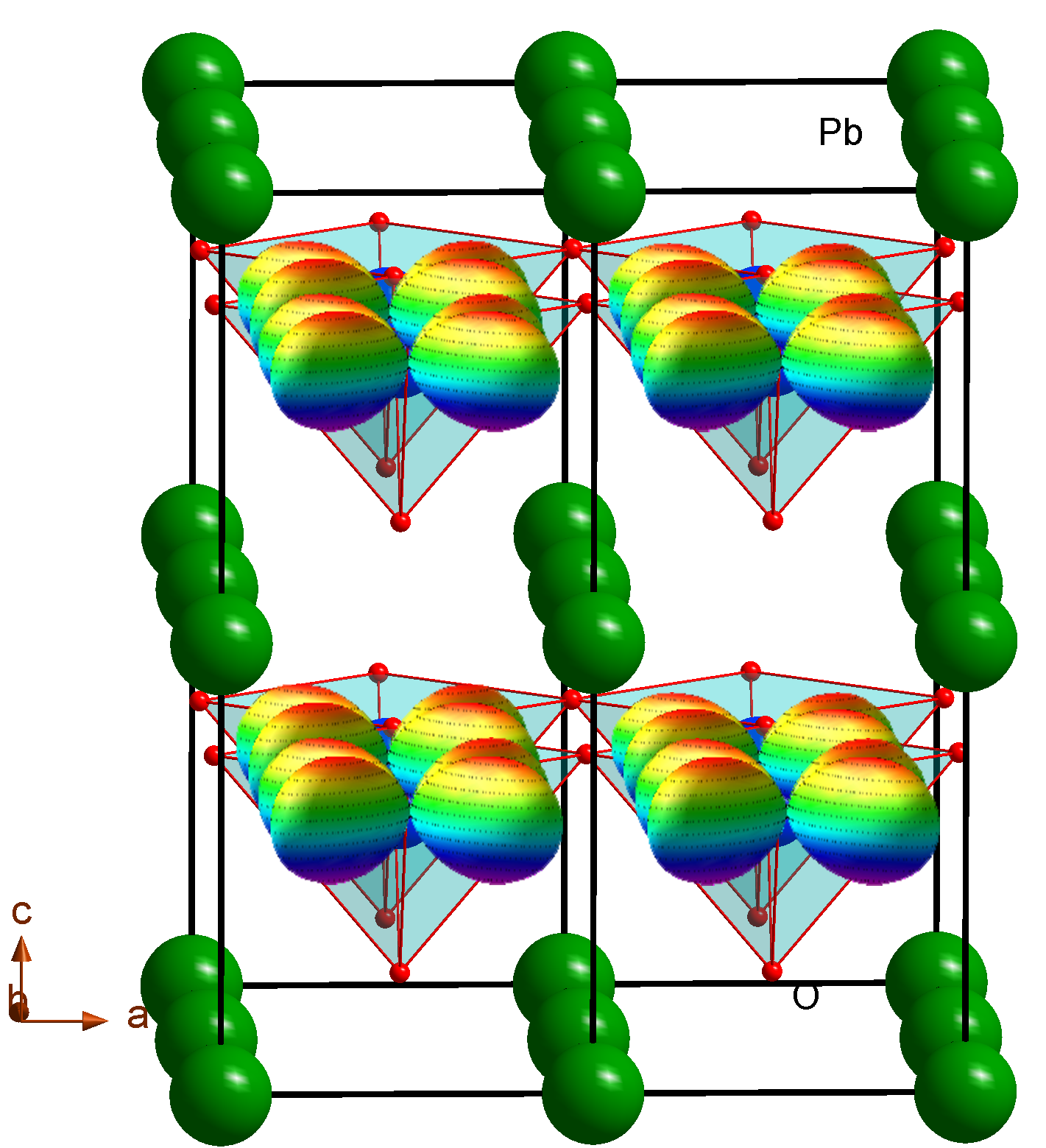}
\includegraphics[width=3in,height=3in]{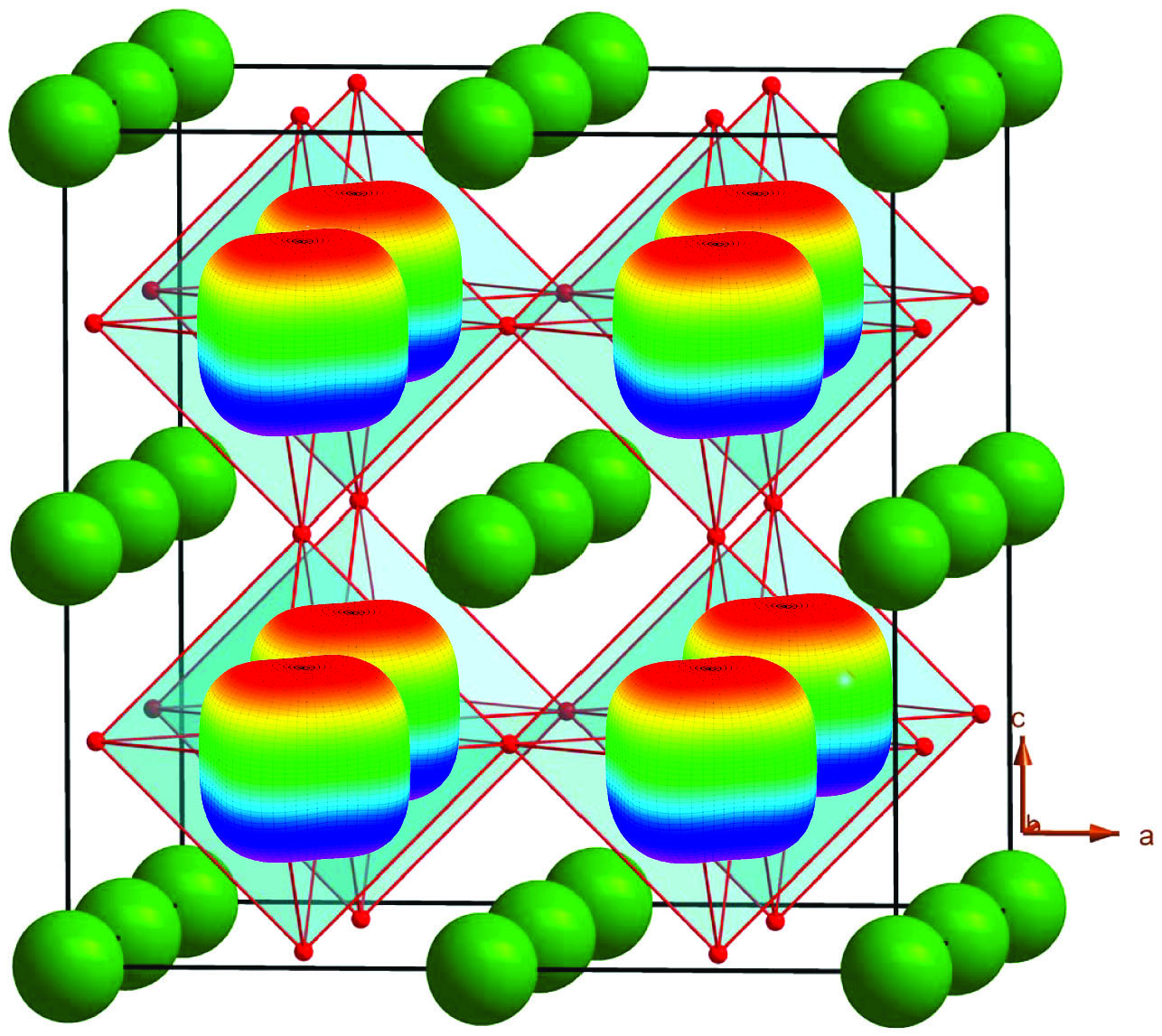}
\caption{Orbital ordering pattern of PVO in (left) the ferroelectric and (right) paraelectric phases as derived from full potential calculations.}
\label{orbord}
\end{figure*}

\par
The orbital projected DOS for the V 3$d$ electrons  for both paraelectric and ferroelectric phases in PVO are given in Fig. \ref{orb} (a) and (b), respectively.  In the presence of a cubic octahedral crystal field the $d$ level of transition metal cation splits into $t_{2g}$ (with degenerate $d_{xy}$, $d_{xy}$ and $d_{zx}$ orbitals) and $e_g$ (with degenerate $d_{x^2-y^2}$ and $d_{z^2}$) orbitals. The $e_g$ levels are completely empty whereas the $t_{2g}$ states are partially filled in the paraelectric phase resulting in a metallic state. The ferroelectric phase of PVO has a square pyramidal coordination (Fig. \ref{afstr}) because of the off$-$center displacement of V. Due to the presence of VO$_5$ square pyramidal environment, the $t_{2g}$ level of V 3$d$ electrons splits into $b_{2g}$ ($d_{xy}$) and doubly degenerate $e_g$ ($d_{yz}$, $d_{xz}$) levels with an energy difference of $\sim$ 1\,eV. Similarly the degeneracy of $e_g$ level breaks to produce separate $b_{1g}$ ($d_{x^2-y^2}$) and $a_{1g}$ ($d_{z^2}$) levels \cite{solovyev2012magnetic}. The orbital degeneracy is broken because of the second-order-Jahn-Teller ordering. This was also noticed in the YTiO$_3$ with $d^1$ configuration of Ti$^{3+}$ similar to V$^{4+}$ in our study\cite{goodenough2015varied}. The orbital projected DOS for the ferroelectric phase given in Fig. \ref{orb} (b) shows that the localized  V 3$d$ electron are present in the lowest lying $d_{xy}$ orbital which is separated from other $d$ orbitals by crystal field splitting, making an spin 1/2 antiferromagnetic ordering in the $ab$ plane. Due to the localized nature of this $d-$electron there is a strong  intra$-$atomic exchange splitting that shifts the unoccupied $d_{xy}$ states to relatively higher energy.

\par
We know that the electrons present in the vicinity of Fermi level actively participate in electrical conduction and magnetic exchange interactions. So we have used the integrated values of the orbital projected DOS i.e. from $E_F$ to -1\,eV to study the orbital ordering. We have used same procedure for OO study as given in our previous studies\cite{vidya2008density,vidya2002spin}, and such analysis not only provides the OO pattern but also a pictorial illustration of the orientation of a particular $d$ orbital. The OO pattern for ferroelectric and paraelectric phases are given in Fig. \ref{orbord}. It can be seen that the V atom has a $d_{xy}$ pattern OO in the ferroelectric phase. This type of orbital ordering in PVO strengthens the intraplanar V-O-V interactions and also stabilizes the 2D magnetism. Hence, the magnetic and orbital ordering are ingeniously coupled with the lattice distortion and ionic displacements that result in strong magnetoelectric coupling. But, in the cubic paraelectric case, the V$-$O$-$V interactions are strong enough to create an itinerant-electron band with all the $t_{2g}$ orbitals partially filled. So, the OO pattern for paraelectric phase does not show particular manner of ordering from any of the $t_{2g}$ orbitals. This weakens the exchange coupling and stabilizes a non-magnetic ground state.

\par
The calculated magnetic moment at the V site is 0.93 $\mu_B$ for the ferroelectric phase. The covalency present between V and O induces a magnetic moment of 0.05 $\mu_B$ at O site. The orbital magnetic moment at the V site evaluated using SO + OP method is estimated to be $-$0.045 $\mu_B$. This indicates that the orbital magnetic moment is aligned antiparalelly with the spin magnetic moment. The magnitude of the orbital moment of PVO is almost 5 times less than that of BiCoO$_3$ \cite{ravindran2008magnetic}. This is due to the the presence of relatively larger spin moment and stronger spin orbit coupling in BiCoO$_3$.

\par
We have calculated the magnetic anisotropy energy\cite{brooks1940ferromagnetic},\cite{bozorth1937directional} for $x$ = 0.5 and 1 in their ground state $C-$AF configurations. For this, we have calculated the variation of total energy by changing the direction of the magnetization with the force theorem for the compositions $x$ = 0.5 and 1. The calculated relative magneto anisotropy energies for $x$ = 0.5 with respect to the easy axis are 0, 0.125, 0.070 and 0.048 meV/f.u. and 0.592, 1.630, 0, and 0.185 meV/f.u. for $x$ = 1 in the [001],[100],[110], and [111] directions, respectively. Therefore, the easy axis for PbTi$_{0.5}$V$_{0.5}$O$_3$ and PVO are [001] and [110], respectively. The change in magnetic easy axis from $x$ = 0.5 to 1 can be attributed to the formation of vanadyl bond with increase in V concentration i.e. the $x$ = 0.5 composition is a BiCoO$_3$ like $C-$AF (so it shows similar easy axis) and with increase in V concentration the tendency to form a 2D antiferromagnetism increases (which also reflects in our band structure).
\par
\begin{figure}[!t]
\includegraphics[scale=0.35]{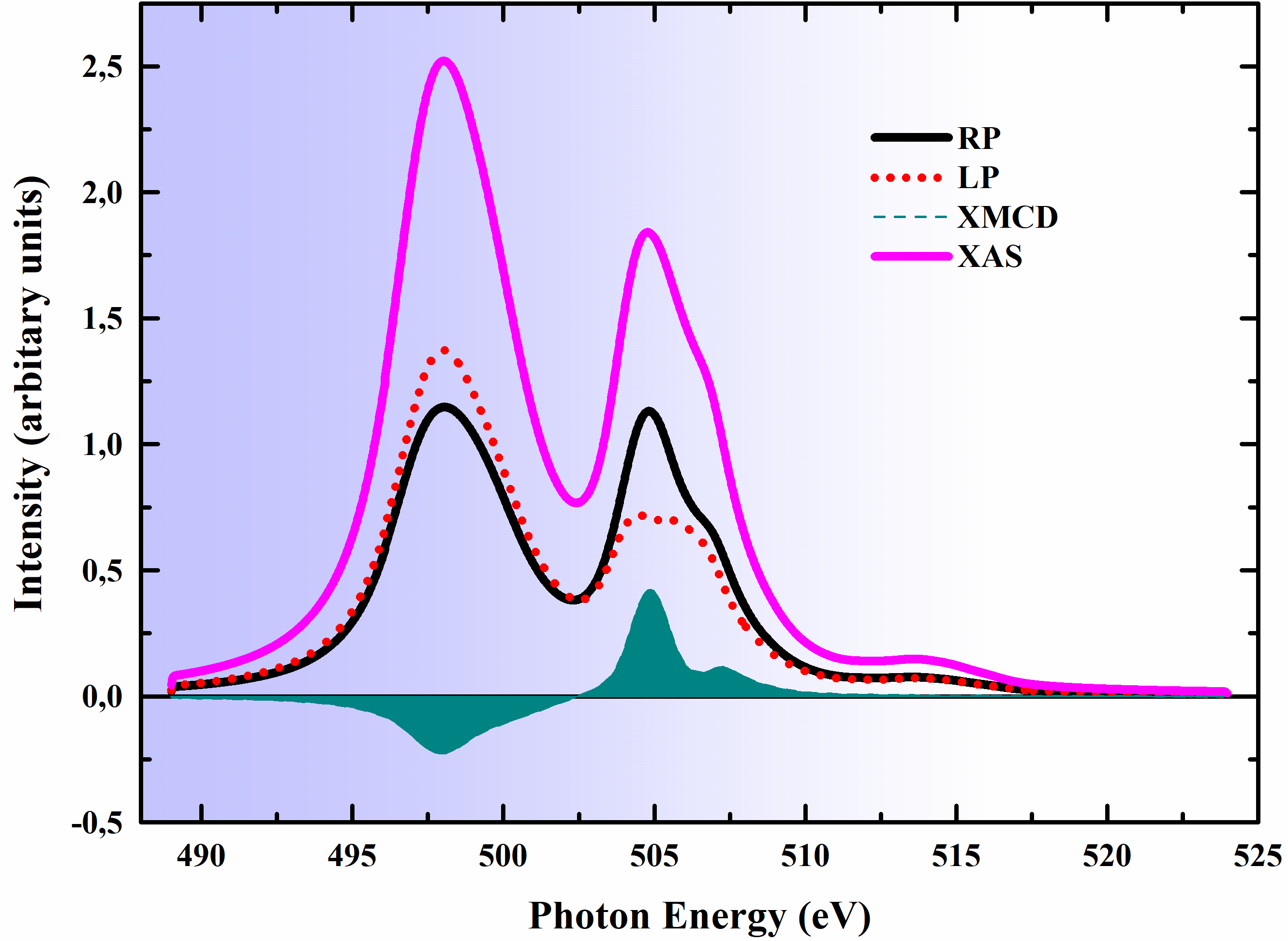}
\caption{Absorption spectra at the V L$_{2;3}$ edges for right circular, left circular polarization of the exciting X-rays, X-ray magnetic circular dichroism and X-ray absorption spectra in PVO}
\label{sumrule}
\end{figure}

In order to get deeper understanding of the magnetic properties of PVO, we have simulated the XMCD and X-ray absorption spectra (XAS) at the $L_{2,3}$ edges of V in the ferromagnetic configuration of PVO. The XMCD and XAS spectra along with the left and right polarized spectra for V atom in PVO are given in Fig.~\ref{sumrule}. By applying the sum rules\cite{carra1993x,thole1992x} given in supporting information, we have evaluated the spin and orbital moments at the V site. The obtained orbital moment of -0.039 $\mu_B$ for V is consistent with that obtained directly from the self consistent calculations (-0.045 $\mu_B$). The sum rule analysis results a spin moment of at 0.8 $\mu_B$/V the V $L_{2,3}$ edges. However, this varies noticeably from the value calculated by SCF method which is $\sim$ 1 $\mu_B$/V. Detailed analysis on different 3$d$ compounds by Wende \textit{et al.}\cite{scherz2004fine} shows that the deviation of the sum rule results from the SCF predictions are larger for lighter 3$d$ elements\cite{wende2004recent,scherz2005limitations,wende2007xmcd} as integral sum rule analysis ignores the spectral shape of the XMCD. So, the detailed fine structure in the XMCD of V should be fitted with the multipole moment analysis as explained in Ref. [67]\cite{wende2007xmcd}.
\begin{table*}
\centering
\caption{Calculated diagonal components of the Born effective charge tensors for $x$ = 0, 0.5 and 1 of PbTi$_{1-x}$V$_x$O$_{3}$ in the ferroelectric phase with ground state magnetic ordering.} 
\renewcommand{\arraystretch}{1}
\begin{tabularx}{\textwidth}{XXXXXXXXXX}
\hline\hline
\textbf{\textit{x~$\rightarrow$}} &  & \textbf{0} &  &  & \textbf{0.5} &  &  & \textbf{1} & \\
\hline
\textbf{Atom} & \textbf{\textit{xx}} & \textbf{\textit{yy}} & \textbf{\textit{zz}} & \textbf{\textit{xx}} & \textbf{\textit{yy}} & \textbf{\textit{zz}} & \textbf{\textit{xx}} & \textbf{\textit{yy}} & \textbf{\textit{zz}}\\
\hline
\textbf{Pb} & 3.641 & 3.641 & 3.331 & 3.623 & 3.617 & 3.664 & 3.251 & 3.232 & 3.275 \\
\textbf{Ti} & 6.166 & 6.166 & 4.961 & 5.480 & 5.406 & 4.778 & $-$   & $-$   & $-$ \\
\textbf{V}  & $-$   & $-$   & $-$   & 4.348 & 4.440 & 4.435 & 4.880 & 4.824 & 3.749 \\
\textbf{O1} &-1.938 &-1.938 &-4.245 &-1.148 &-1.154 &-3.098 &-1.442 &-1.525 &-3.720 \\
\textbf{O2} &-2.603 &-5.266 &-2.024 &-2.078 &-3.875 &-1.417 &-2.209 &-4.029 &-1.764 \\
\textbf{O3} &-5.266 &-2.603 &-2.024 &-3.716 &-1.899 &-1.423 &-3.877 &-2.513 &-1.724 \\ \hline\hline
\end{tabularx}
\label{BEC_table}
\end{table*}


\par
\subsection{Born effective charge and Spontaneous Polarization}
\label{sec:charge}
\par
We have calculated BECs for all the compositions mentioned above. The diagonal components of the BECs for $x$ = 0, 0.5 and 1.0 compositions are given in Table~\ref{BEC_table}. The formal valence for Pb, V, Ti and O in PbTi$_{1-x}$V$_x$O$_{3}$ are +2, +4, +4 and -2, respectively. But, the diagonal components of the calculated BEC tensors for the atoms are much larger than their nominal ionic values. This reveals that there is large dynamic contribution superimposed to the static charge. This is due to the strong covalency effect where a large amount of non-rigid delocalized charge flows across this compound during the ionic displacement. This results in additional charges with respect to the nominal ionic values (anomalous contribution) which are obtained at the atomic sites. A large BEC value at the oxygen site also suggests that a large force can be generated by applying a small field which can stabilize the polarized state. 
\begin{figure}[!b]
\includegraphics[scale=0.45]{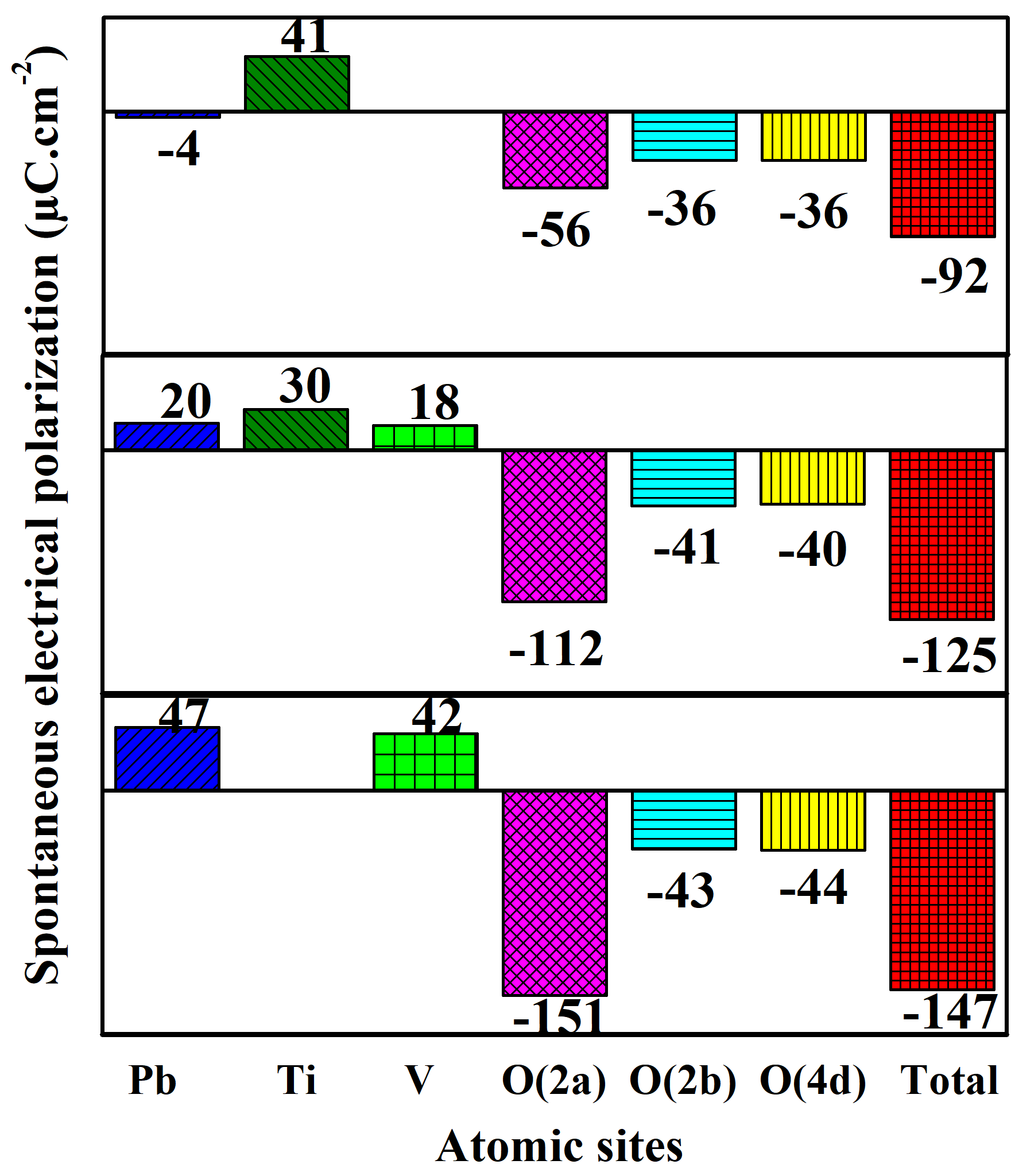}
\caption{Calculated site projected spontaneous electric polarization of PbTi$_{1-x}$V$_x$O$_{3}$ for $x$ = 0 (upper panel), 0.5 (middle panel) and 1 (lower panel).} 
\label{fig:ppol}
\end{figure}
In an effort to shed light on the polarization of PbTi$_{1-x}$V$_x$O$_{3}$ and its origin, we have evaluated the polarization of PbTi$_{1-x}$V$_x$O$_{3}$ for $x$ = 0 to 1. PTO and PVO are well studied ferroelectrics with polarization values of 92 $\mu$C/cm$^{2}$ and 147 $\mu$C/cm$^{2}$, respectively. The calculated polarization values increase from 92 $\mu$C/cm$^{2}$ for pure PTO to 105, 110, 118, 130, 135, and 147 $\mu$C/cm$^{2}$ for $x$ = 0.25, 0.33, 0.5, 0.67, 0.75, and 1, respectively. This is due to the increase in tetragonal distortion with $x$ as discussed above. The polarization calculated with point charges showed comparatively low values. The reduction in polarization values using point charges indicates that covalency effect plays an important role deciding the polarization of these materials.
\par
The atom$-$decomposed spontaneous polarization is given for $x$ = 0, 0.5 and 1 compositions in Fig~\ref{fig:ppol}. The major contribution towards the polarization is coming from the displacement of the O atom from its centrosymmetric position. It can be seen that the polarization contribution of Ti towards higher $x$ value decreases. In contrast, an opposite trend is seen for V atom. This is because of the enhancement of covalency between B site cation and O with V substitution at Ti site.

\subsection{Conclusion}
PbTi$_{1-x}$V$_x$O$_{3}$ possesses strong coupling between magnetic and electric order parameters for V concentration $>$ 12.3\%. Lower substituted compositions are non-magnetic ferroelectrics. Also, from the other side, the inclusion of Ti at the V site increases the chance of forming a 3D arrangement of V cations rather than 2D. The strong covalency between Ti/V$-$O and noticeable covalency between Pb$-$O bring the non$-$centrosymmetric distortion and stabilize the ferroelectric ground state. The ferroelectric-to-paraelectric transition is accompanied by a magnetic to non$-$magnetic transition for $x$ $>$ 0.123. There is a large volume contraction during the above transition, indicating a strong lattice$-$ferroelectric coupling and strong magnetovolume effect. The calculations show high values of spontaneous electrical polarizations which are mainly due to the displacement of the apical oxygen in the BO$_6$ octahedra in the paraelectric phase due to strong Ti/V$-$O covalent interaction. Therefore, the PbTi$_{1-x}$V$_x$O$_{3}$ is a multifunctional series of compounds having giant magnetoelectric effect where one can change the magnetic properties drastically (even from magnetic to non$-$magnetic as shown here) by applying electric field and vice versa, strong magnetovolume effect, induce 2D magnetism as well as orbital ordering by electric field, strong ferrolectric$-$lattice coupling etc. Also, recent experimental studies by Zhao Pan \textit{et al.}\cite{pan2017colossal} show that for lower V concentration, this series shows good NTE response and also a temperature induced tetragonal to cubic transition happens which makes this series more interesting for the researchers to investigate its magneto/electro/multi$-$caloric properties. With all these properties coexisting in the series, PbTi$_{1-x}$V$_x$O$_{3}$ can be useful for many multifunctional device applications.
\label{sec:con}
\section{Acknowledgements}
The authors are grateful to the Research Council of Norway for providing computer time (under the project number NN2875k) at the Norwegian supercomputer consortium (NOTUR). This research was supported by the Indo$-$Norwegian Cooperative Program (INCP) via Grant No.F. 58-12/2014(IC). L. P. wishes to thank Prof. Anja Olafsen Sj{\aa}stad for her fruitful discussions. L. P. also thanks Mr Ashwin Kishore MR and Dr. A. Krishnamoorthy for critical reading of the manuscript. 

\section{Associated content}
Energy vs. c/a ratio curves, partial dos for $x$ = 0.50 with GGA+$U$ method, total DOS for $x$ = 0.33, 0.50, and 0.67 with GGA+$U$ method, the Briilouine zone path used for band calculations, electronic band structure for $x$ = 0.50 with GGA+$U$ method, electronic band structure for $x$ = 1 with GGA+$U$ method, variation of energy difference between magnetic and nonmagnetic states with $x$, and sum rule analysis. This information is available free of charge via the Internet at http://pubs.acs.org. 
\section{Author Information}
\subsection{Corresponding Author}
*Email: raviphy@cutn.ac.in \newline
Tel: +91 94890 54267
\subsection{Notes}
The authors declare no competing financial interest.
\bibliography{PTVO}
\end{document}